\documentclass[journal,onecolumn]{IEEEtran} 
\usepackage{graphicx}

\usepackage[utf8]{inputenc}
\usepackage{amsmath}

\usepackage[T1]{fontenc} 

\usepackage{hyperref} 

\usepackage{listings}

\usepackage{mathtools}
\usepackage{graphicx}
\usepackage{tabularx}
\usepackage{subfigure}
\usepackage[table]{xcolor}
\usepackage{dblfloatfix}
\usepackage{fixltx2e}
\usepackage{multirow}
\usepackage[justification=centering]{caption}

\usepackage{tabularx}					
\usepackage{array}						
\usepackage{booktabs}					

\usepackage{multirow}
\usepackage[table]{xcolor}

\begin{document}

\title{Integrating micro-learning content in traditional e-learning platforms}


\author{Rebeca P. D\'iaz-Redondo
       Manuel Caeiro-Rodr\'iguez
       Juan Jos\'e L\'opez-Escobar
       Ana Fern\'andez-Vilas
\thanks{atlanTTic Research Center, School of Telecommunications Engineering, University of Vigo, Spain \\
rebeca@det.uvigo.es, mcaeiro@det.uvigo.es, juanjo@det.uvigo.es, avilas@det.uvigo.es}
}




\maketitle

\begin{abstract}
\noindent Lifelong learning requires appropriate solutions, especially for corporate training. Workers usually have difficulty combining training and their normal work. In this context, micro-learning emerges as a suitable solution, since it is based on breaking down new concepts into small fragments or pills of content, which can be consumed in short periods of time. The purpose of this paper is twofold. First, we offer an updated overview of the research on this training paradigm, as well as the different technologies leading to potential commercial solutions. Second, we introduce a proposal to add micro-learning content to more formal distance learning environments (traditional Learning Management Systems or LMS), with the aim of taking advantage of both learning philosophies. Our approach is based on a Service-Oriented Architecture (SOA) that is deployed in the cloud. In order to ensure the full integration of the micro-learning approach in traditional LMSs, we have used two well-known standards in the distance learning field: LTI (Learning Tools Interoperability) and LIS (Learning Information Service). The combination of these two technologies allows the exchange of data with the LMS to monitor the student's activity and results. Finally, we have collected the opinion of lectures from different countries in order to know their thoughts about the potential of this new approach in higher education, obtaining positive feedback.
\end{abstract}



\section{Introduction}
\label{sec:introduction}

Continuous learning has become an increasing need in our society: the constant and rapid evolution of knowledge requires workers to adapt to this new context in order to maintain their productivity. Micro-learning comes into play to facilitate this process to potential learners by breaking down new concepts into small fragments or pills of content, also called micro-content. These small learning units are given to learners progressively and in a way that is suited to them. 

The combination of several factors has stimulated the development and the positive reception of the micro-learning paradigm. First, the human capacity to stay focused on a single item, avoiding distraction and inattention, has decreased. Indeed, some very specific studies have revealed that users only pay 8 consecutive seconds of continuous attention when surfing the Internet \cite{Hayles2007}. Second, very quick changes in all areas, especially in technology, have resulted in workers needing to update their training constantly. Some studies estimate that workers who lack the appropriate IT skills waste 21\% of their working time \cite{Webster2012}. Finally, traditional training is not proving to be a good method to train workers effectively and efficiently, since: (a) long training processes are required, so work is interrupted; (b) long, laborious creation processes are needed -it is estimated that 43 to 185 hours are needed to generate the content for a single hour of training, even more for more specialized training \cite{talent18}- and, even less promising, (c) the impact of the method is very limited. In fact, several studies have concluded that only 15\% of workers are able to adequately apply the knowledge they have acquired and that 80\% of such knowledge will be forgotten after just one month. 

As a result, traditional formal education has progressively been relegated to the background, representing less than 6\% nowadays, while 74\% of the time devoted to in-company training courses already integrates distance-learning environments based on new technologies or combines both approaches in mixed learning environments (blended-learning) \cite{Zhang2011}. In a corporate environment, this mixed strategy must be understood not only as a combination of formal and face-to-face pedagogical approaches with non-formal ones, but also as a combination of all of them with the training that is intrinsic to the work experience. One of the most widely accepted mixed models is 70/20/10, where 70\% of the training comes from work experience, 20\% corresponds to the content taught by instructors or mentors and the remaining 10\% is related to formal training. This strategy has proven to be so effective that it has progressively been adopted by different companies in their training processes \cite{Margaryan2004} \cite{Merrill2009}. Usually, traditional e-learning platforms, as LMSs, are used in the corporative sector, specially in SMEs (Small and Medium-sized Enterprises). 

In any case, offering adequate environments that can be totally integrated into the working processes is not an easy task, which usually results in high drop-out rates. This is clearly reflected in the use of MOOCs (Massive Open Online Courses) as a complement to daily work activity, with more than 90\% of drop-outs in this context, due to a range of reasons that can be summarized in four big blocks \cite{Eriksson2017}: (i) content (difficulty or unsuitability); (ii) design of the course; (iii) personal situation of the learner (limited knowledge of the language, difficulty to access the Internet, etc.), and (iv) difficulty to combine this activity with other daily tasks. The last reason is the most frequent one. On the contrary, a more optimistic scenario is shown in \cite{Stracke2017}, where it is questioned whether these drop-out rates are a good way of measuring the success or failure of MOOCs. This study proposes an interesting premise: simply, trainers and trainees may not have identical objectives. A throughout analysis has shown that content accessibility is the main motivation of many students, who use the platform at their convenience for training without following the guidelines of the course in an exact way. In any case, and while still trying to determine if drop-out rates should be considered a reliable indicator of failure or success, it is quite certain that they are high enough to be considered symptomatic of a lack of suitability of the offered training to meet the needs of students. 

Anyway, the difficulty in designing training content that is adequate for learners is not restricted to the field of continuous training for adults/employees, but it can also be used in other areas such as the university. This problem is so relevant that it is included in the strategic agendas of most European countries. Indeed, it is treated with a very high priority in more than half of them \cite{EU2015}. However, the method used to assess success or failure is based on different parameters, such as the completion rate, the retention rate or the time-to-degree. Besides, these parameters are calculated under different basis. In such a diverse and wide context, we can find all kinds of educational approaches which, with the aid of new digital technologies, try to strengthen not only the learner’s motivation but also the learning processes themselves in order to facilitate their integration in the learner’s social and training contexts: (i) those whose main occupation is studying and (ii) those that have to combine work and study (corporate learning).

Despite the fact that different proposals have been defined in an attempt to solve these problems, the present document will focus just on one of them: micro-learning, a didactic technique that has arisen in this context with the aim of facilitating the training process to students through the decomposition of new concepts in small fragments or pills, also known as micro-content. A micro-content focuses exclusively on the transmission of relevant information, with the help of visual, interactive and very brief elements, so the student is less likely to drop out or tune out during training. These small lesson plans are gradually given to students in a way that is tailored to their needs, with the aid of telematics devices. In the specific case of corporate training, micro-learning seems to be one of the most appropriate mechanisms, at least according to some studies \cite{Anil2012}. For those employees who must be trained as an indispensable process in their professional career, micro-learning perfectly fits in by introducing short units of content, usually in audiovisual format, which can be digested in downtimes between activities \cite{Zhang2011}. This paradigm, however, does not only benefit trainees, but also their trainers. Since the content has to be clear and concise, it usually requires from less resources and time in the authoring process. Thus, this allows for greater agility when generating new content tailored to the rapidly changing needs of corporate environments.

This paper is organized in two parts. The first one provides a detailed revision of the state-of-the-art of the micro-learning paradigm and a detailed description of its main characteristics and good practices for designing micro-learning activities and content. It also includes a summary of different approaches in the literature to provide architectures and frameworks for micro-learning, like the use of mobile devices, Service-Oriented Architectures and Cloud Computing. Finally, some successful commercial tools are also described. In the second part of the paper, we introduce our approach for integrating the micro-learning paradigm in more traditional e-learning platforms. Although micro-learning has been proposed in the literature as a stand-alone solution, we consider that it should also be considered as a good mechanism to reinforce the traditional Learning Management Systems (LMS) platforms. This hybrid approach combines the advantages of the micro-learning paradigm and traditional solutions. To the best of our knowledge, there is no other proposals that try to combine these two fields. Both the student and the trainer can benefit from this combination: the former can enjoy the attractive and easy-to-consume micro-content, while the latter can use both methods in the same platform. Our solution is based on a Service-Oriented Architecture (SOA) of different services that are deployed in the cloud. In order to guarantee the full integration of the micro-content in a traditional LMS, we propose using Learning Tools Interoperability (LTI) and Learning Information Service (LIS) as the two standards that enable the micro-content to exchange data with the LMS and register the student's feedback on their profile. In short, our approach allows trainers to bring together the benefits of traditional learning or blended-learning and the new opportunities that the micro-learning paradigm has to offer. Finally, we have also conducted a survey to collect the opinion of lectures from different countries about the potentiality of micro-learning. The results show that lectures consider this approach can be used in their lessons to improve the knowledge transmission and students engagement and it constitutes a very good supplement for traditional blended-learning.

The structure of the paper is the following. Sect. \ref{sect:MicrolearningParadigm} focuses on defining the micro-learning paradigm and its characteristics. Sect. \ref{sect:OverviewMicrolearning} overviews the most relevant approaches can be discerned in the literature. Sect. \ref{sec:proposal} introduces our proposal to fully integrate micro-content in an LMS, defining micro-content and the architecture and detailing the technicalities of the solution. In Sect. \ref{sec:evaluation}, we describe the survey that we conducted among international lectures in order to know their opinion about the potentiality of this new learning approach and in Sect. \ref{sec:discussion} we discuss about advantages and disadvantages of the micro-learning paradigm. Finally, Sect. \ref{sec:conclusion} summarizes our approach and future work.

\section{Micro-learning as a training paradigm}
\label{sect:MicrolearningParadigm}

Although there are all kinds of definitions of micro-learning, none of them has been unanimously accepted. Theo Hug’s is perhaps the most widely assumed, which is based on seven dimensions \cite{Hug2006}: 
\begin{itemize}
\item {\em Time}. A limited effort that leads to short time requirements.
\item {\em Content}. Short units with well-delimited subject matters and relatively simple problems.
\item {\em Curriculum}. Parts of modules or parts of curricular content, brief didactic elements, etc.
\item {\em Format}. Diversity of formats, such as fragments, pills, lab assignments, etc.  
\item {\em Process}. Activities that are either independent or integrated into a wider context, iterative processes, etc.
\item {\em Media}. Classroom-based learning or distance learning based on different multimedia content.
\item {\em Learning models}. Repetitive, reflexive, pragmatic, constructivist, concept-based, connectivist, etc.
\end{itemize} 

This reflects the main feature which defines micro-learning: its capacity for integrating a huge variety of didactic parameters without restricting or limiting any of the options in the seven identified areas. While the concept of micro-learning cannot be considered novel, since it dates back to the sixties, it was not until the arrival of the Web 2.0 age that this concept reappeared strongly, with the term being finally coined in 2004. In fact, the Web 2.0 context was the ideal breeding ground for this kind of didactic approach. The {\em do it yourself} philosophy that lies behind the concept of prosumer (consumer and producer of content) and the ease with which new technologies allow users to create and add new content have resulted in a higher tendency to use micro-formats: simple, brief and very specific content such as podcasts, blogspots, wiki-pages; or short messages on networks, such as Twitter or Facebook.
   
Furthermore, the social facet so characteristic of these tools has allowed, on the one hand, micro-learning to spread quickly and, on the other hand, the learning task itself to be reinforced due to the creation of user communities that share interests and have the support they need to interact with others as well as with content \cite{Buchem2010}  

Mobility is also very important in micro-learning. Dealing with reduced content specifically designed to be easily shared and/or added implicitly leads to the fact that these units, weakly coupled, can be easily transported or integrated into different contexts and devices.   

Finally, micro-learning is closely linked to the concept of micro-training: micro-tasks that can be either used autonomously or integrated into mixed training contexts (blended-learning), which represents an interesting and efficient option (in terms of time and cost) for corporate environments.

The combination of the previously-mentioned features (brevity, concision, quick creation, aggregation and distribution) makes it possible for micro-content to be easily integrated into the daily activities and selected depending on the particular needs and/or interests of each learner. As a result, even though this strategy can be successfully used in many contexts, it is particularly adequate for those professionals who need to update their knowledge while they continue with their work activity.  


In short, the idea of micro-learning can be considered opposite to the concept of macro-learning: a training context which is also based on new technologies, but where training is more extensive, could be more formal and is based on the most common technology platforms, such as LMSs. However, they must not be regarded as two irreconcilable alternatives, but rather as two different techniques that provide different solutions to the particular needs and purposes of each user, which does not mean they cannot be combined as necessary. In fact, our the approach we introduce in this paper goes in line with this hybrid philosophy by combining both strategies. Thus, we propose to offer short self-contain learning pills (typical from micro-learning) that are more oriented to informal consumption as a supplement of traditional learning content (such as documents, activities, etc.) thought for a formal monitoring. Usually, this traditional content (i) demand trainers to monitor the students activity; (ii) demand students to invest more time and (iii) demand from LMSs to support the interaction between trainers and trainees. 

Therefore, micro-learning can be defined as a technique that allows for distance training, but provided in small amounts that the learner can assimilate in brief training periods which are interspersed with other activities. For this approach to be successful, it is essential to pay attention to the design of micro-content as well as how it is sequenced, that is, to the design of micro-learning activities. 

\subsection{Design of micro-learning activities and micro-content}
\label{sect:DesignOfMicrocontent}

In order to support a micro-learning approach, it is necessary to provide activities and to facilitate the design of adequate training sequences. Regarding the {\em micro-learning strategies} the related literature offers a wide range of options: self-directed learning \cite{Knowles1975}, situated learning \cite{Lave1991} and community-based learning \cite{Wenger1998}. For co-creating and sharing content, it is also worth mentioning collaborative learning and process-oriented learning \cite{Davenport2004}. The literature also recommends that the {\em micro-learning processes} be designed combining micro-learning sessions (micro-content) of no more than 15 minutes each which can be organized in weakly coupled cycles of three sessions: introduction, activity and conclusions. The {\em micro-learning activities} should be designed to be directly managed by learners. Moreover, these activities are expected to facilitate active participation by promoting the exploration, the use and the creation of content (e.g. elaboration of mind maps, production of visual content, interactive activities, etc.). The {\em Microlearning materials} should be able to draw attention to very specific and clear aspects; besides, there should be complementary materials which enable learners to directly participate in their generation, assembling and modification. It is important to find a balance between brief format and complementary information. Finally, it is especially interesting to promote the {\em micro-learning for communities}, where materials derived from the training activities themselves are accessible for the student community in a manner that they can serve as a basis for debate, as support material for new activities or simply as a reference to consult. \\

Taking these guidelines into account, the next step is designing micro-content, for which it is necessary to observe the following principles \cite{Leene2006} \cite{Linder2006}: 
\begin{itemize}
\item {\em Format}. Units should be designed in a way that they are brief and easily perceived at a glance (e.g. without resorting to scrolling down) as well as lightweight enough to be rapidly distributed among different environments (e.g. simple structure, low resolution).
\item {\em Focus}. Objectives and topics must be clear and easily expressible in a few brief and concise sentences.  
\item {\em Autonomy}. Each piece of micro-content must be independent so that learners do not have to search for additional information.
\item {\em Structure}. They should condense simple, though necessary, information (title, topic, authors, date, labels, etc.). 
\item {\em Simple access}. Micro-content should be designed to be hosted as a single resource on the Internet, but should also be easily accessible from any other location.
\end{itemize}
    
Even if micro-content is appropriately designed and sequenced, it is not suitable for all learning contexts. In fact, it is not adequate if the concepts to be acquired are complex. Micro-learning activities are more convenient to supplement the acquisition of skills that are strengthened through repetition and practice. In addition, the interleaving of micro-learning activities can become misleading for those learners who believe that they are able to do multiple tasks at once, which is not the objective. Indeed, it should be noted that completing all the activities of a micro-lesson does not directly lead to learning the concepts to be acquired, as occurs with all other teaching techniques \cite{Jomah2016}    

\subsection{Duration of micro-learning activities}
\label{sect:MicrolearningDuration}

There is not a widely accepted criterion for the establishment of the most appropriate duration of micro-learning sessions. As previously stated, it is thought that they should not exceed 15 minutes. In any case, different studies have been conducted to analyze this so as to keep the students’ attention by using diverse information sources.  Perhaps the most in-depth analyses are those undertaken in the visual context (viewing of videos), as this format is the most widespread both in non-formal educational environments (MOOCs) and in the transmission of information in general (advertisements, news, etc.). It is interesting to note that, although micro-learning activities go further than watching a video, the results of the following analysis about the viewer attention give a clue about how to proceed when designing micro-learning activities.

\begin{figure}[htbp]
  \centering
  \includegraphics[width=0.6\textwidth]{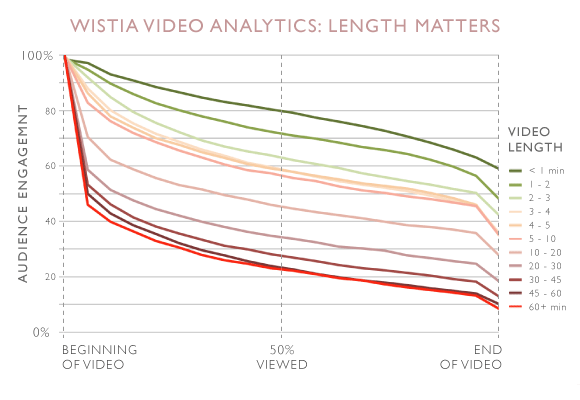}
  \caption{Relation between number of views and video length [{\small \tt https://wistia.com/blog/does-length-matter-it-does-for-video-2k12-edition}]}
  \label{fig:wistia}
\end{figure}

A first analysis, using data taken from platforms such as YouTube, reveals that the most successful videos are those whose length is within the range from 60 to 90 seconds. In fact, a study carried out by WISTIA (a video-hosting platform) shows that the percentage of views decreases as video length grows (Fig. \ref{fig:wistia}).   

Nevertheless, it should be noted that this conclusion mainly applies to video ads, whose objective is being the least invasive possible. Indeed, a deeper analysis focused on videos of eminently disseminative and didactic content, reached different conclusions: the most successful average duration for podcasts is 22 minutes, the average duration of a TED talk is 18 minutes and most of the video-seminars with the greatest impact (webinars) last approximately 10 minutes.

\begin{figure}[htbp]
  \centering
  \includegraphics[width=0.6\textwidth]{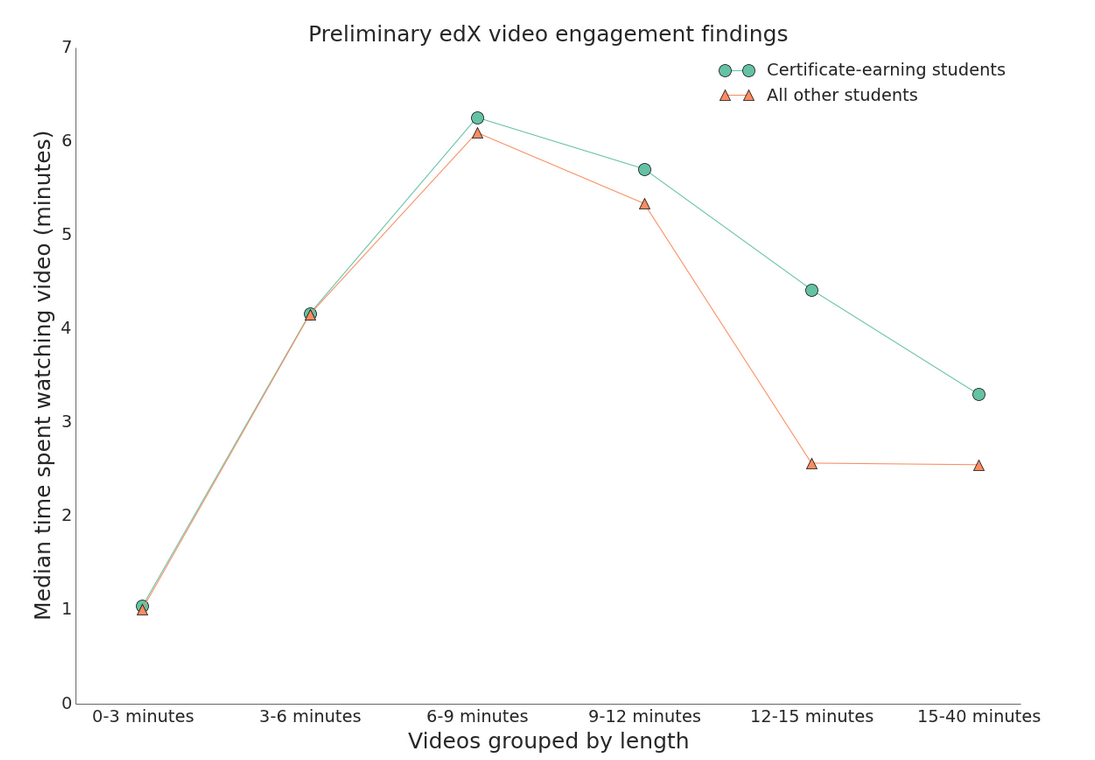}
  \caption{Analysis of the most appropriate duration for educational content [taken from \cite{Guo2014}]}
  \label{fig:guo}
\end{figure}

If we focus only on the length of the videos used in MOOCs, \cite{Guo2014} collected data from more than 6 million views from a total of $862$ videos which are available in the edX platform for $127,839$ students. It can be concluded that those videos which are over 9 minutes in length produce a drop in attention, which leads to the fact that students start skipping parts of the video, resulting in a net viewing of around 20\% of the total. In this article, videos are recommended to last between 6 and 9 minutes in order to capture the students’ attention and to keep it during the entire video, so that the maximum efficiency of the resource can be achieved (Fig. \ref{fig:guo}).

\begin{figure}[htbp]
  \centering
  \includegraphics[width=\textwidth]{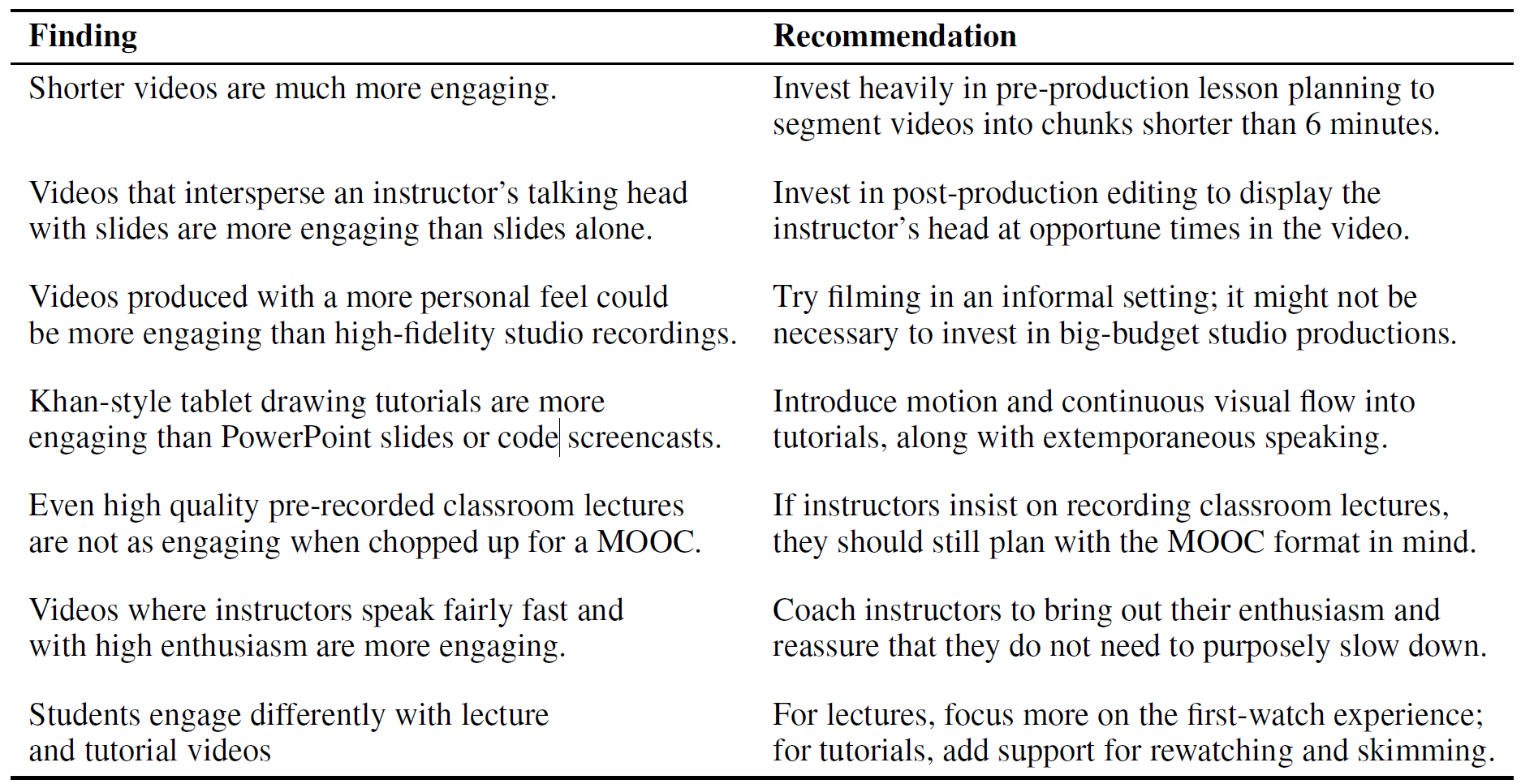}
  \caption{Design of videos for micro-learning activities [taken from \cite{Guo2014}]}
  \label{fig:guo2}
\end{figure}

Moreover, the following items (contained in Fig.\ref{fig:guo2}) were analyzed: 
\begin{itemize}
\item In order to keep the viewer’s attention, it is more effective to combine images of the instructor while speaking with images of the learning content (typically, slides and screenshots), rather than resorting only to the latter. Additionally, recordings where visual contact is prioritized achieve better results. 
\item {\em Khan style} videos are those where the instructor is typically using a tablet to give outlines or brief notes that accompany the speech. This approach has proved effective. It is recommended, to the extent possible, to include short annotations in the visual material (codes or slides).
\item Pre-production allows for better results, since it is based on a deeper analysis of the design, the sequencing and the content of the video. Post-production cannot solve content-related problems where pre-production has not been as adequate as it should have. 
\item Despite the fact that some studies \cite{Williams1998} state that 160 words per minute are the optimum rate to capture the audience attention, it should be noted that, in the present study, even greater rates have been found to be effective. For this reason, it is advisable to reduce pauses in post-production if necessary. 
\item Finally, videos for learning purposes are usually classified into two large groups: those where theoretical concepts are explained (declarative) and those where more practical aspects are addressed (tutorials). The analysis carried out found that students only watch between 2 and 3 minutes of the tutorials (regardless of their length), even though they normally watch them on more occasions than declarative videos, skipping parts of the video more frequently. Consequently, it is advisable to offer learners the possibility of skipping certain parts of the tutorials in a guided way so that it is easier for them to find what they are looking for
\end{itemize}

\section{Overview of micro-learning approaches}
\label{sect:OverviewMicrolearning}

\subsection{Integrated Micro-Learning}
\label{sect:integratedMicroLearning}

One of the first approaches was created for computers and it used the screensaver of the devices to detect the inactivity of the users (when switching on, when accessing a remote server, when accessing a website, etc.) and trigger training activities. A first version, defined in \cite{Gassler2004}, was coined as Integrated Micro Learning (IML), where the integration of the training activities in daily life was carried out in a manner that regular activity could only be resumed after having accomplished a training micro-activity. In order to assess the level of acceptance of this system, a small study where users could refuse micro-training was carried out. However, after the pilot phase, it was concluded that 75\% of the activities were accepted. Later, in 2007, another pilot study was conducted with 74 cards and 30 users. Results showed that users consumed 15 cards per day on average (in other words, 3,000 cards per year if considering that a year has a total of 200 working days), a very positive figure with regard to the potential acceptance of this training paradigm.      

The evolution of this system allowed for a more elaborate deployment over Windows XP that made it possible to use more sophisticated cards such as multiple choice-only one correct or multiple choice-multiple answers. Training activities were initiated after finishing a task but before starting a new one, so the system was not exclusively focused on screensaver activation. The first pilot test to evaluate this deployment was conducted in 2010 with 62 users, who were public workers of the Austrian government. In this case, besides monitoring the records provided by the tool itself, a satisfaction survey was carried out. Results were very positive, as 91\% of the users that had begun the course (82\%) succeeded to complete the training, with an average consumption of 143 cards. The satisfaction survey endorsed these data since users found the experience very positive.  

\subsection{Micro Mobile Learning}
\label{sect:microMobileLearning}

Although micro-learning techniques can be used with all kinds of devices, it should be noted that portable devices (telephones, tablets, etc.) are clearly the most convenient for this new paradigm. It is easy to deduce that both paradigms go hand in hand if understanding mobile learning as {\em "Any sort of learning that happens when the learner is not at a fixed, predetermined location, or learning that happens when the learner takes advantage of the learning opportunities offered by mobile technologies"} \cite{OMalley2005}. In fact, some definitions of micro-learning explicitly mention the possibility of resorting to on-the-move learning: {\em "Microlearning is a new research area aimed at exploring new ways of responding to the growing need for lifelong learning or learning on demand of members of our society, such as knowledge workers. It is based on the idea of developing small chunks of learning content and flexible technologies that can enable learners to access them more easily in specific moments and conditions of the day, for example during time breaks or while on the move."} \cite{Gabrielli2005}.   

Both micro-learning and mobile learning are based on reusable, autonomous, self-sufficient and linkable content. Micro-content is usually adequate for both paradigms; thus, some authors deal with micro-content design without taking into account the paradigm for which it will be employed \cite{Souza2014}. As a consequence of these similarities between both approaches, the term Micro Mobile Learning has already been mentioned in the literature to refer to the conjunction or integration of both, being this conjunction characterized by brevity, ubiquity and interactivity \cite{Bruck2012}. Brevity, because micro-content is designed in brief blocks; ubiquity, because the training can be carried out anywhere; interactivity, because the interaction between the trainer and the student is crucial in the process. 

Because of these reasons, the technological evolution led the work started in \cite{Gassler2004}, described in the previous section, to a third deployment based on the use of mobile devices, co-financed by two European projects. The first pilot assessment test was carried out in 2011, with the participation of 22 public workers from different sectors of the United Arab Emirates government who were presented with a training sequence of 71 cards. In this case, users were free to use the client for mobile phones or personal computers (PC). In addition to the good monitoring results of the training, users showed their satisfaction in a follow-up survey. Nowadays, this system is distributed to different companies and administrations.

\cite{Zhao2010} addresses another aspect: the use of Micro Mobile Learning for professor training purposes. These authors identified three different models: (i) independent learning, where learners decide their own learning path (goals, how to tackle different problems, etc.); (ii) question-based learning, where trainers ask different questions to learners so that they can work towards their resolution, and (iii) collaborative learning which establishes a common aim for a group of learners who have to work together in order to achieve it. Transversely, three different interaction modes are included in each of these models: Mode 1 refers to the usage of a short message service (SMS), Mode 2 refers to an interactive consultation mode and Mode 3 refers to the use of web browsers.   

\begin{figure}[htbp]
  \centering
  \includegraphics[width=\textwidth]{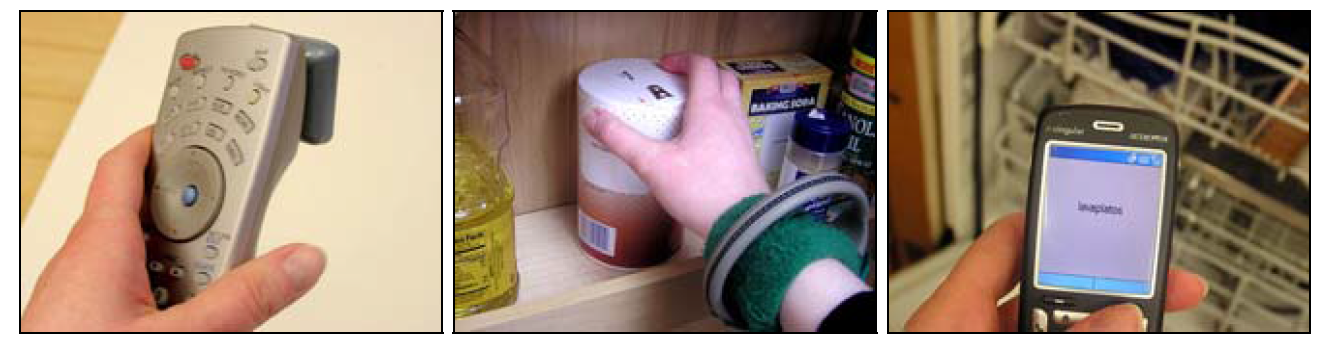}
  \caption{Different sensors used to trigger a micro-learning session [taken from \cite{Beaudin2007}]}
  \label{fig:MITExperiment}
\end{figure}

A completely different alternative was proposed by \cite{Beaudin2007}. In this case, interaction is not directed by the instructor and there does not exist collaborative interaction between the trainees either, regardless of the students' needs, i.e. this approach fits better as a ubiquitous IoT (Internet of the Things). In this context, the learning process is led by the trainee's context. These researchers, who belong to the House\_n research department of the MIT ({\small \tt http://web.mit.edu/cron/group/house\_n/}), propose to place a set of sensors throughout the learner's house to enable interaction with the learner's mobile phone in order to trigger micro-content. When these sensors detect the smartphone and/or specific activity of the users, a micro-learning session is triggered. All kinds of sensors are employed (Fig. \ref{fig:MITExperiment}): from cards, buttons and/or RFID stickers, to movement sensors adhered to the remote control of the television. To test this system, the authors decided that the objective would be to learn new vocabulary in a foreign language (Spanish). The system was tested by a married couple for four weeks, during which time words and phrases were presented to the users with an average frequency of 57 per hour. Despite these figures, the users found the interaction interesting and effective. 

\begin{figure}[htbp]
  \centering
  \includegraphics[width=0.5\textwidth]{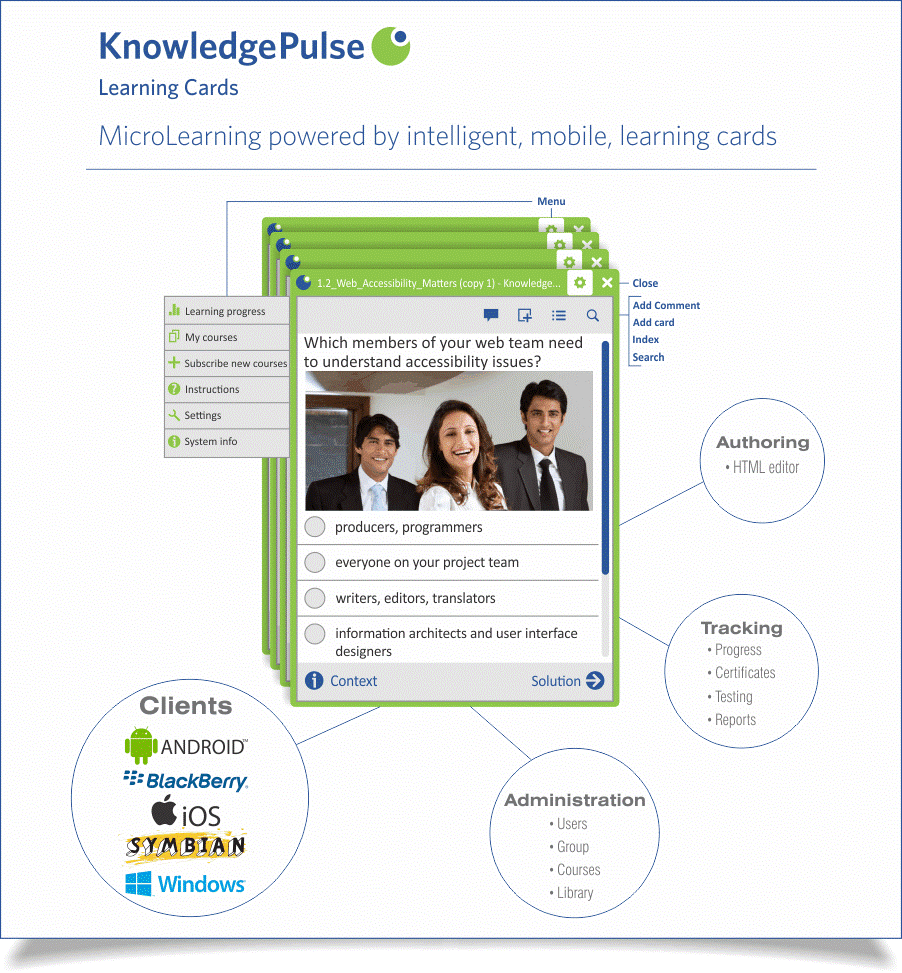}
  \caption{Commercial image of KnowledgePulse Microlearning [{\small \tt http://www.videotelephony.com/knowledge-pulse-micro-learning/}]}
  \label{fig:knowledgePulse}
\end{figure}

The best example of success in this area is probably the one described in \cite{Bruck2012}, where a micro-learning environment specifically oriented to the use of mobile devices is proposed. This environment uses the KnowledgePulse® MicroLearning system (KP\footnote{{\small \tt http://www.videotelephony.com/knowledge-pulse-micro-learning/}}), developed by Research Studios Austria FG (RSA FG)\footnote{{\tt https://www.researchstudio.at/en}}, to which the authors of the publication adhere. Fig. \ref{fig:knowledgePulse} shows the commercial image of the product. This system sends small units of educational content in the form of interactive learning cards, being the content particularly suited for mobile devices such as mobile phones. Learners are asked to give a short answer, normally just true or false. Continued interaction with content generates a workflow that strengthens the acquisition of new knowledge. This learning environment is more oriented to the memorization of propositional content and uses a learning algorithm \cite{Bruck2005} \cite{Bruck2008} which is customized depending on the answers and the interaction of each student. Subsequently, the system, in its version KnowledgePulse® 2.0, is also capable of supporting certain social interaction and collaboration. Thus, users can create their own cards and share them with peers.

This system is based on the Leitner system, named after the German journalist Sebastian Leitner, who devised it in the seventies. This procedure consists of the spaced repetition of short texts that need to be memorized. These texts are organized in groups or boxes. When the student acquires the new knowledge, the text changes boxes or groups, in a way that it will be presented again at a longer interval. When the knowledge has not been acquired correctly, the text goes to another box or group and is presented to the trainee again at a shorter interval. 

The solution provided by KnowledgePulse® MicroLearning is based on this method. The trainer has to previously divide the content into a set of learning cards, in a manner that each of them represents a different step in the memorization process. These steps will need to be sequenced according to a didactic sequence. Taking all this into account, the KP generates the repetition processes needed to guarantee the memorization of the content in the short and long terms. Besides, the KP system detects when the device (mobile phone) is not being used in order to activate a training card, the one corresponding to the user, which will depend on personal track record and on the didactic sequence of knowledge.   


\subsection{Composition of micro-content: an SOA approach}
\label{sect:micro-learningSoa}

The composition of micro-content to elaborate or define training sequences requires a system which can provide solutions to store, to locate and to compose micro-content. The design of the solution given by \cite{Arroyo2005} is based on the SOA (Service Oriented Architecture) paradigm, where communications are managed through an ESB (Enterprise Service Bus). This flexible proposal undoubtedly raises a structure in layers where the definition of enhanced metadata for the description and characterization of micro-content is particularly important. The lower layer, or Persistence layer, is intended to store the elements to be combined, that is, the micro-content (Learning Objects, LO), with the support of the metadata set that describes them. The following layer, the Semantic Layer, is made up of two broad modules that are intended to manage micro-content so that it is semantically compatible with the requirements of the Semantic Web Services (SWS). The following layer, the Semantic Web Services Layer, is intended to manage the web services which publish, discover, negotiate and combine micro-content. Finally, the authors also provide a tool to annotate the content in a simple way.          

Ontological support for the description of educational content (WSMO extension) was conceived as an extension of the WSMO ontology (Web Service Modeling Ontology) \cite{W3C2005}, which was in turn extended to adapt to the specific requirements of micro-content by using, for this purpose, the LOM (Learning Object Metadata, \cite{IEEE2002}) standard. Two main difficulties were faced. First, the representation of learners to ensure that they can be characterized through their own competences and learning styles. Second, the interconnection of educational elements through a range of relations: different versions, authoring, technical requirements, etc. In fact, these relations make it possible to establish an educational sequence through the composition or sequencing of micro-content \cite{Cobo2006}.

\subsection{Micro-learning content in the cloud}
\label{sect:micro-learningCloud}

The cloud computing paradigm fits in perfectly with the concept of micro-learning, since if the latter is characterized by the use of brief educational resources upon request, the cloud computing paradigm is also supported by the flexible use of resources (hardware, software, storage, computation, etc.) whose amount dynamically varies depending on what is needed at each particular moment. Consequently, the application of this paradigm to the provision of micro-learning environments can be considered as a natural evolution from the first platforms based on web environments, which would allow the availability of storage, backup services and computer services, elastically provided depending on the needs at each moment, and at a more affordable cost. In addition to the typical advantages of this paradigm, also applicable to most of the software and hardware solutions (cloudonomics), micro-learning systems could benefit from (i) the creation of libraries of educational micro-resources that are accessible in a simple and functional way; (ii) the use of micro e-portfolios, where the progress made by each user can be registered; (iii) the customized access to micro-resources depending on these micro e-portfolios and (iv) easier real-time interaction with other users, by using micro-blogs, micro-chats and different systems of intercommunication based on the cloud computing paradigm \cite{Li2011}.

In \cite{Kovachev2011} an architecture characterized by the following elements is proposed: (i) use of portable devices (smartphones and tablets), since they facilitate mobility, dynamism and flexibility, all of which are inherent to micro-learning; (ii) application of OCR (Optical Character Recognition) programs to process the data entries, since the use of such programs makes this task faster and simpler, thus more convenient for the learner, and (iii) application of web scraping techniques to quickly analyze the content of web pages and highlight the most relevant information. The combination of these three elements allows for a workflow where users, after accessing educational content, use software embedded in their browser (add-on) in order to add different elements of various web pages with the aid of a web scraper. Finally, this enhanced content is updated in the cloud and can be synchronized to any portable devices to access the micro-learning educational experience whenever the user deems appropriate. 

Finally, cloud computing allows enhanced content to be subsequently accessed by other users, which results in a collaborative and dynamic environment that is inherent to micro-learning. With this aim, the aforementioned authors \cite{Kovachev2011} show a typical micro-learning cycle (introduction, activity and conclusions) and provide a labelling system based on six categories, context, frequency, recentness, semantics, preference and feedback, which help to classify and retrieve micro-learning content.

In the same vein, the research work of \cite{Sun2015} proposed a service-oriented system which enables the organization of a Virtual Learning Environment (VLE) for portable devices (telephones, tablets, etc.) based on collaborative learning and micro-learning. The architecture of such VLE consists of three functional modules that have been respectively implemented in order to: (i) obtain historical information about the learner (track record); (ii) capture data of the learner while interacting with the system and (iii) synchronize with the cloud. Additionally, as the most innovative contribution, the VLE incorporates two services: TaaS (Teamwork as a Service) and MLaaS (Micro-learning as a Service).

The first service, TaaS \cite{Sun2014}, deployed in Amazon EC2, combines the benefits of collaborative work with the implementation of the learning flow. The module has five web services (Survey Service, Jigsaw Service, Bulletin Service, Monitor Service and Inference Service) which should preferably be kept active and working in sequence so that all benefits can be obtained, even though they can be decoupled and adapted to the interests of the VLE. The Survey Service provides the necessary support to guarantee communications from mobile devices despite the potential instability of connections when using these devices. The Jigsaw Service makes it possible to organize discussion between learners, whereas the Bulletin Service allows assigning tasks to different work teams. As for the Monitor Service, it facilitates that all the students can monitor the work of the rest of the team members. Finally, the Inference Service is intended to assign specific tasks to each student depending on individual learning styles and preferences.

\begin{figure}[htbp]
  \centering
  \includegraphics[width=\textwidth]{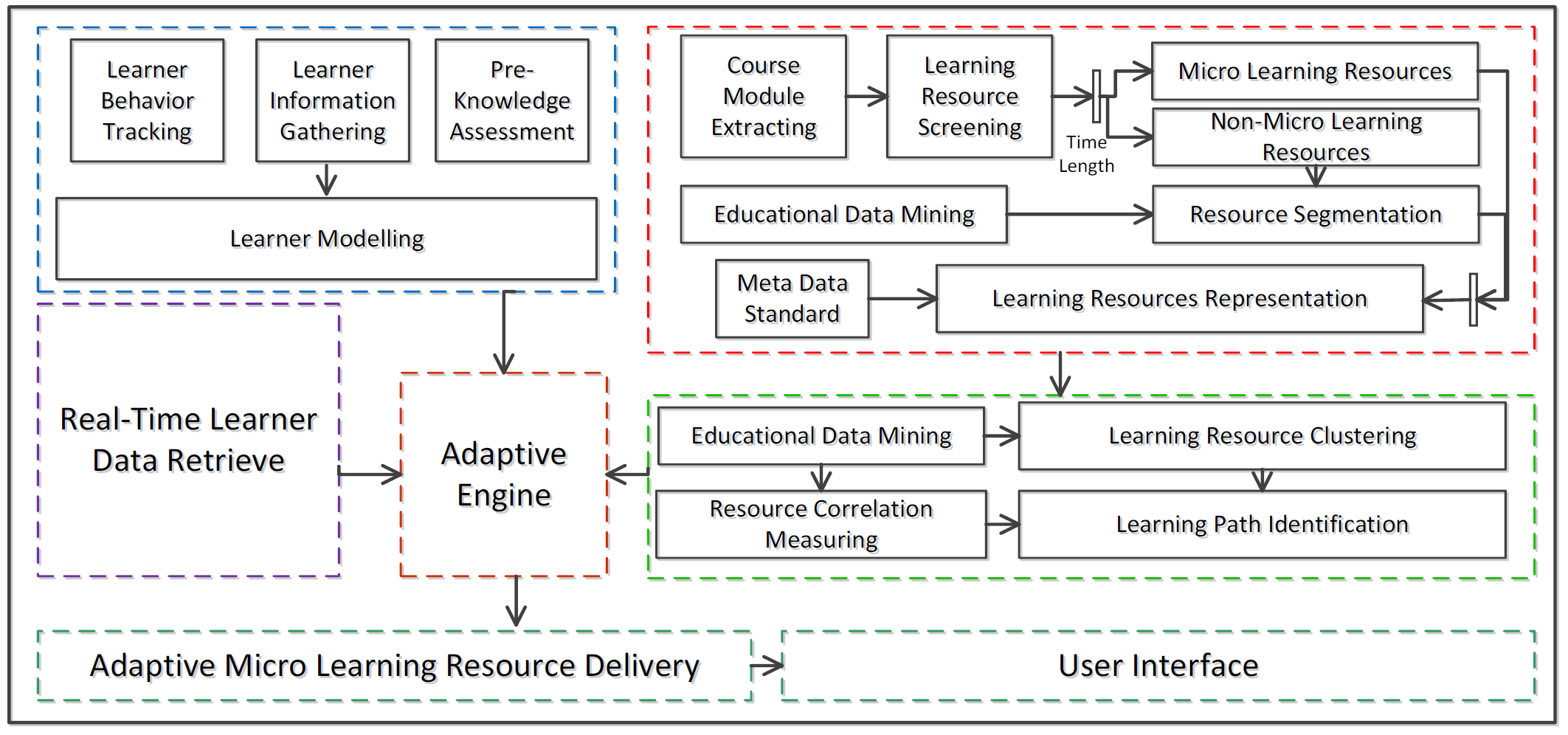}
  \caption{MLaaS architecture [taken from \cite{Sun2015}]}
  \label{fig:MLaaS}
\end{figure}

The second service (MLaaS, also implemented in Amazon EC2) is intended to provide micro-content for those users who prefer to resort to their mobile devices to take advantage of little spare moments (e.g. while commuting to and from work). The architecture shown in Fig. \ref{fig:MLaaS} was designed with this aim. Here, the Learner Modelling block is dedicated to characterize the learner (track record, interaction with the system), while the Real-Time Learner Data Retrieve Service is intended to obtain data on the progress and the availability of learners (in other words, when they prefer to study and the amount of time they devote to this activity). The Learning Resources Representation block is intended to manage all the possible representations of the available micro-content so that the most adequate version can be provided as needed. For this purpose, micro-content is labelled by using standard metadata for the description of micro-units \cite{Beydoun2009} \cite{Veeramachaneni2013}, by incorporating information such as keywords, duration, language, popularity, difficulty, etc. Finally, the Adaptive Engine uses the information that comes from these three sources or sub-modules to adapt content to each learner’s learning styles and preferences, learning context and individual needs. Therefore, this engine becomes the centre or core of the global system: it is in charge of selecting and adapting the content which is most appropriate for each micro-learning process.          

This is an inclusive proposal of micro-learning in an environment based on the cloud, oriented to mobile devices and underpinned by collaborative work. The same authors continued their work by analyzing a very relevant issue when dealing with adaptive environments: cold-start problems, that is, problems to begin the adaptation when enough information to characterize the students is not available. While it is a recurrent problem in all the adaptation or recommendation environments, \cite{Sun2017} suggested an adequate solution to the OER (Open Education Resources) that is also integrated into the global proposal as a module oriented to service and is provided in the same architecture in the cloud. The full study, with all the intermediate solutions, can be consulted in      
\cite{Sun20172}.

\subsection{Commercial approaches}
\label{sect:MicrolearningSuccess}

There is a wide range of options when deploying a micro-learning platform. In spite of the difficulties they must face, some proposals have proved successful in this context.
The most noticeable ones, which perhaps cannot be directly associated with this paradigm, are YouTube, TED and the Kahn Academy. YouTube allows users to organize their own training when they desire to acquire some new skill or knowledge in a fast and effective way (for example, to prepare a dish with the aid of a new recipe, to meet a new origami challenge, to install a new device, etc.). Indeed, YouTube has become a source of multidisciplinary and ubiquitous knowledge, comparable with Wikipedia in terms of dissemination but used for the gradual display of content in a training context, with a very enjoyable and digestible format: video. The second of the above-mentioned proposals, TED, is based on the altruist dissemination of knowledge spread by international or local experts (in its TEDx version) through brief recorded talks that can be easily watched via its web platform. Finally, the Khan Academy is a non-profit educational organization founded by Salman Khan, who is a graduate of the MIT and the Harvard University whose main objective is to provide structured knowledge in the form of brief lessons or presentations on certain topics that the learner can freely select on the web platform. Moreover, the Khan Academy enables learners to access interactive content in order to put in practice the knowledge they have acquired.                       
In addition to these three proposals, it is worth mentioning some other approaches:
\begin{itemize}
\item The KnowledgePulse® MicroLearning system \\ 
{\small \tt http://www.videotelephony.com/knowledge-pulse-micro-learning/}, which was developed by Research Studios Austria FG (RSA FG\footnote{{\tt https://www.researchstudio.at/en}}, as described in section 5. The system is aimed at a very specific public: trainees in corporate contexts.
\item Grovo ({\small \tt https://www.grovo.com/}), a company founded in 2010, has offered a solution based on the SaaS (Software-as-a-Service) paradigm since 2013. This proposal allows for combining and adjusting micro-content: short videos (144 seconds), brief audio lessons and short and interactive visual elements. Trainees can combine these multi-format elements to enhance their learning experience. Since this solution is intended for corporate environments, it allows trainers and managers to use activity-monitoring tools based on modules of data analytics. Fig. \ref{fig:Grovo} shows the appearance of the platform when accessed by the manager and the trainees. It has been successfully employed in large international corporations such as PepsiCo or Gap Inc.

\begin{figure}[htbp]
  \centering
  \includegraphics[width=0.75\textwidth]{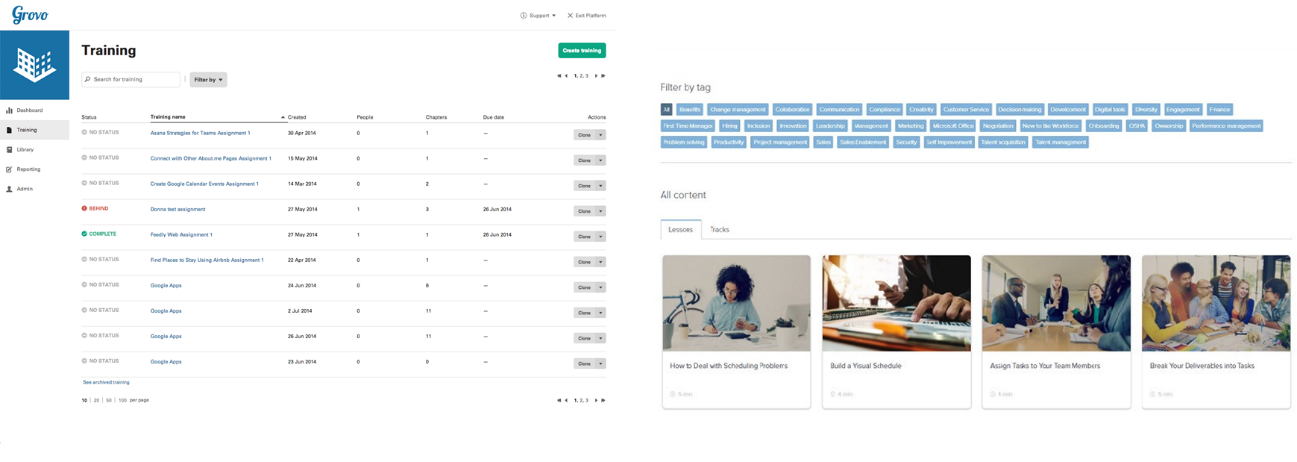}
  \caption{Appareance of the Grovo platform for managers and trainees [{\small \tt https://www.grovo.com/}]}
  \label{fig:Grovo}
\end{figure}

\item The Coursmos platform ({\small \tt http://coursmos.com/}), founded in 2013, also provides micro-content. They are normally brief lessons that take no more than three minutes, which make up over 38,600 courses that usually take no more than fifteen minutes in total. This platform was originally developed for mobile devices, especially telephones, this being the reason why it was distributed through the Apple Store and Google Play. Nevertheless, a web solution was launched in 2014. It includes a module of recommended content for learners, who are more than 1,700,000 nowadays. Additionally, Coursmos incorporates a clearly collaborative orientation allowing its users to create content. As shown in Fig. \ref{fig:Coursmos}, there are four different Access modes, from the most basic one to those specifically intended for enterprise environments. Each of them offers a particular range of functionalities, which is reflected in their cost. Differences are mainly in terms of storage capacity, number of courses that are supported, number of events, number of authors, the possibility of customizing the domain, use of monitoring tools and applications adapted to mobile devices, etc. Nowadays, the platform is being successfully used for training purposes in more than 100 companies. 

\begin{figure}[htbp]
  \centering
  \includegraphics[width=0.75\textwidth]{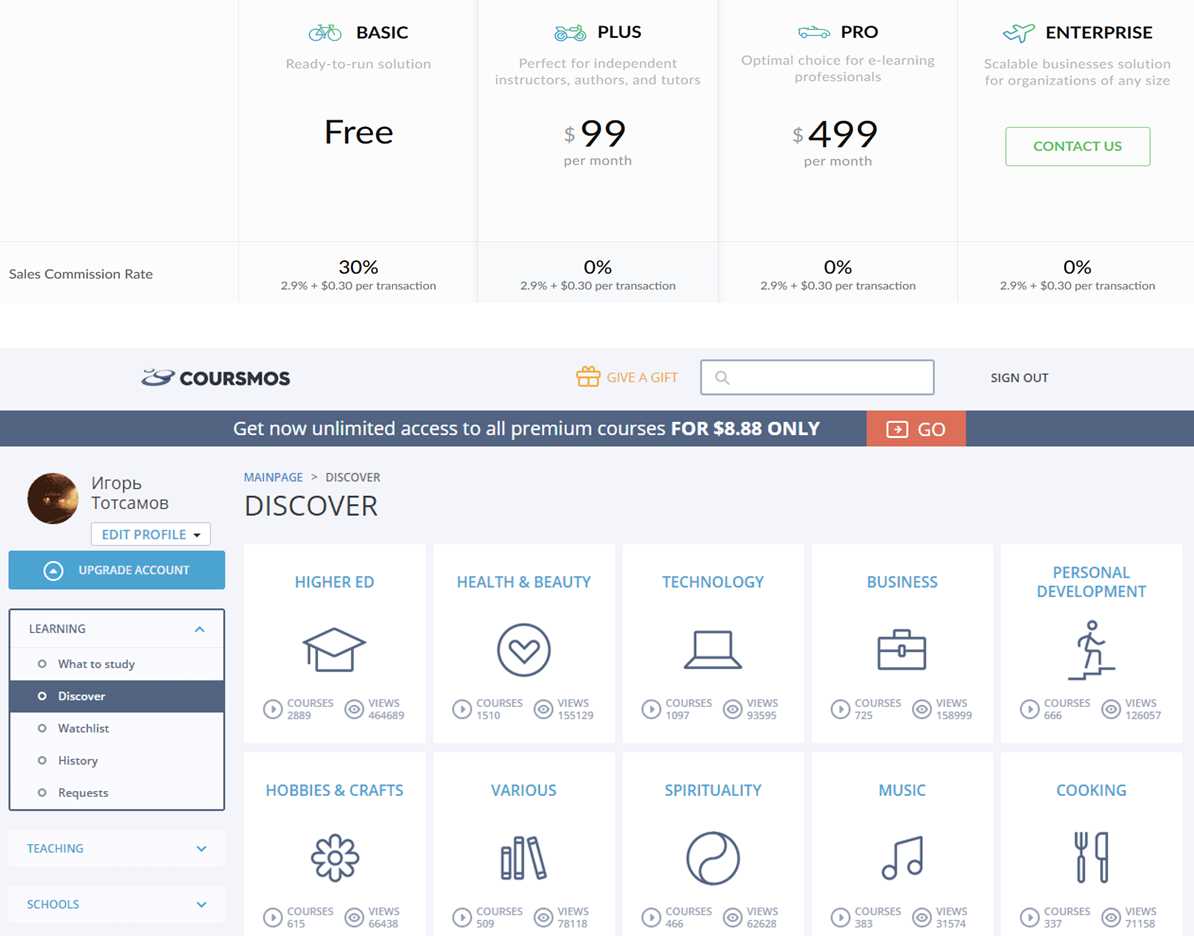}
  \caption{Deployment modalities of the Coursmos platform and its student interface [{\small \tt http://coursmos.com/}]}
  \label{fig:Coursmos}
\end{figure}

\item There are different solutions for mobile devices that propose micro-training environments, even though this term is not explicitly used. Among them, it is worth mentioning Duolingo ({\small \tt https://www.duolingo.com/}), which is a language learning app. Students subscribe to the language they want to learn and the app gradually provides them with sequences of flashcards, in a similar way to that described in (Bruck et al., 2012), as explained in section 5 of the present document. As learners progress through the course, the flashcards that are presented to them become more and more complex. Phrases, vocabulary and grammar are transmitted through the audio and the images, without resorting to videos. Moreover, the tool has an audio recognition system to identify whether the student’s pronunciation is adequate or not. Fig. \ref{fig:Duolingo} shows the appearance of the application: an example of the progress made by a learner in the different headings can be seen on the left, whereas one of the tests that the student has to pass to keep making progress is shown on the right. This application incorporates gamification and socialization techniques so that the students can publish their progress in the social media and receive mentions or hallmarks, which entitle them to extra time to use the application for free.      

\begin{figure}[htbp]
  \centering
  \includegraphics[width=0.75\textwidth]{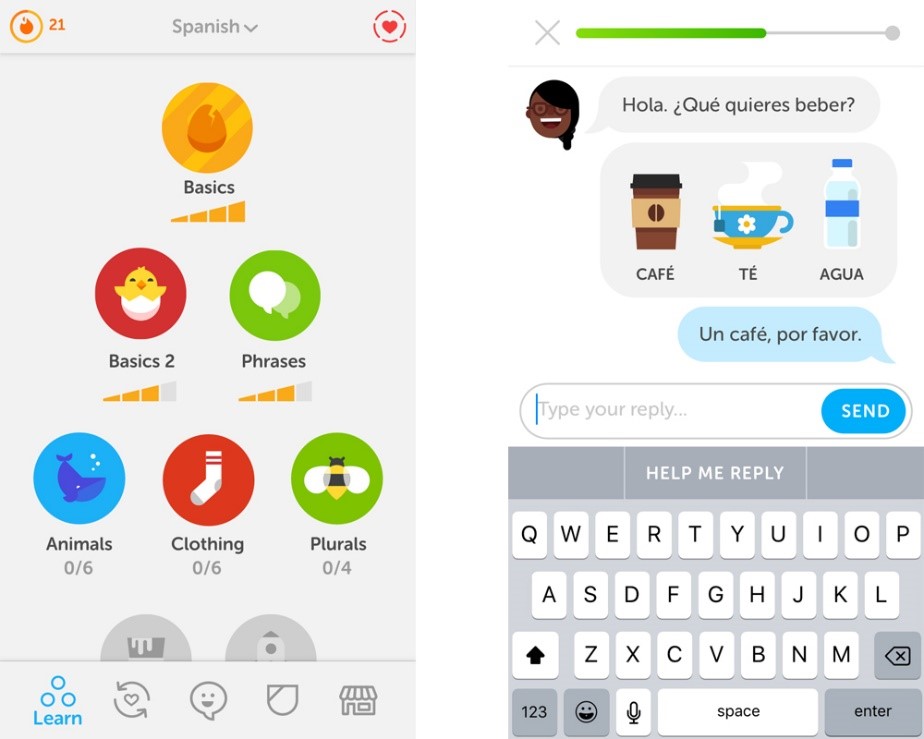}
  \caption{Appearance of Duolingo and its corresponding assessment test [{\small \tt https://www.duolingo.com/}]}
  \label{fig:Duolingo}

\end{figure}

\end{itemize}

\section{A hybrid approach: integrating micro-learning in a traditional e-learning solution}
\label{sec:proposal}


As mentioned above, continuous learning is currently an increasing need due to the constant and rapid evolution of knowledge that forces people, especially workers to quickly update their skills and knowledge. However, formal education is not considered an appropriate option for the corporate context. Not only that, a hybrid approach that combines formal training, work experience and informal training has proven to be effective and is being progressively adopted by different companies \cite{Margaryan2004} \cite{Merrill2009}. An integration of micro-learning activities in traditional distance learning frameworks is precisely what we propose. This hybrid solution merges the advantages of the paradigm of micro-learning and the advantages of the LMSs.

First, and in spite of the existence of some autonomous solutions for student management and interaction with micro-content, we propose to take an existing learning platform as a starting point. It is advisable to use: (i) an LMS, such as Moodle ({\small \tt https://moodle.org}), widely used in education and with a modular structure that enables the integration of new elements; or (ii) a platform created specifically for non-formal education and capable of managing a high number of students, such as OpenEdX ({\small \tt https://open.edx.org/}), probably one of the most popular platforms for MOOC environments. The advantages of using an existing learning platform include (i) familiarity with the environment on the part of both technicians and professors, (ii) a range of tools to manage users, profiles, permissions, documents, etc. and (iii) continuous updates and improvements in terms of security and functionality. 

Second, micro-content should be created, hosted and managed with the aid of another platform to enable, for instance, collaborative creation and social interaction. With this aim, it is advisable to use a {\em Service-Oriented Architecture} (SOA), where micro-content is exchanged among the different components of the software that manages it to offer functionalities such as classification, search, composition and/or modification. All these services form the micro-content management platform. It is highly recommended to implement it in a cloud computing environment, since this combines the advantages of both paradigms: the loose coupling of an SOA and the elasticity of cloud computing.

\begin{figure}[htbp]
  \centering
  \includegraphics[width=0.8 \textwidth]{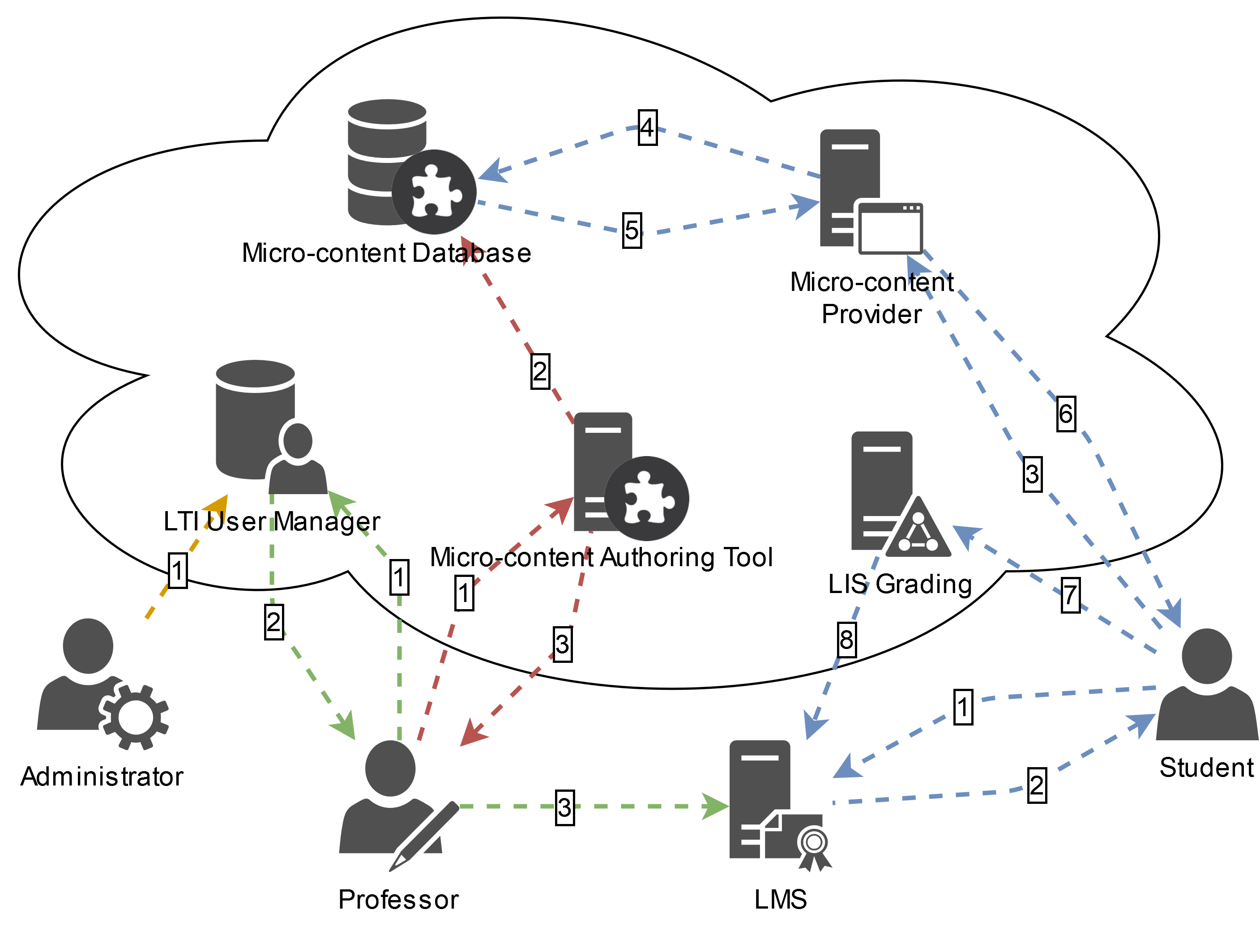}
  \caption{Cloud architecture for the micro-learning platform}
  \label{fig:cloud-platform}
\end{figure}

Our approach is depicted in Fig. \ref{fig:cloud-platform}: the upper part of the figure shows the different SOA services that are hosted in a cloud computing platform. The lower part shows the three different roles that interact with the system (administrator, professor and student) and the LMS where the micro-learning content is going to be displayed and available for the students. Using an LMS (i) allows us to include micro-learning content as simple references or URIs to be displayed in the LMS, and it (ii) does not require the development of specific applications for students to access the learning content. 

The proposed SOA (Fig. \ref{fig:cloud-platform}) is accessible by several user profiles according to their rights and actions. The flow in red represents the interaction between the Authoring Tool and the Professor: (1) a professor registered in the Authoring Tool creates the micro-learning content via a web platform, (2) which is stored through the HTTP API of the Micro-content Database, and (3) an URL is given to correctly deploy the micro-content in the LMS. The flow in blue represents the interaction between the student and the LMS: (1) the Student gets access to the LMS and (2) works with the available learning contents; other interactions are totally transparent to the Student. The flow in yellow shows the interaction between the Administrator and the micro-learning platform: (1) the Administrator grants the Professor the authorization they need to upload micro-learning content to the LMS.

One of the main requirements to integrate micro-learning content into an LMS is that this external (interactive) content should be able to exchange data with the LMS. For this to be possible we analyzed different alternatives before finally defining our proposal. The following sections will explain our approach in detail. First, in Sect. \ref{sec:proposal-authoring} we introduce the authoring tool that produces micro-content according to the main guidelines in the literature. Second, in Sect. \ref{sec:proposal-lti} we deal with the problem of integrating micro-learning content in an LMS and finally, in Sect. \ref{sec:proposal-lti-details}, we detail the technicalities of our integrated solution. 

\subsection{Micro-content and its authoring tool}
\label{sec:proposal-authoring}

According to the main recommendations found in the literature, we consider that each unit of micro-content must be composed of three elements: a brief introduction or description, the development or explanation of the content and a small section to assess knowledge acquisition. Each piece of micro-content and the set of metadata that describes it will be an indivisible unit. Preferably, the development or explanation will be audiovisual; should this not be possible, it will be either interactive or visual. As for the assessment section, it should consist of short-answer or multiple-choice questions.

Although it is not the main of this paper, it is worthy to note that we have also developed an authoring tool (one of the SOA services depicted in Fig. \ref{fig:cloud-platform}) to assist professors when creating micro-content. Although there are a lot of pieces of software to create e-learning resources, they are intended for ad-hoc applications. That is the reason why we have decided to develop our own authoring tool oriented to micro-content that we can upgrade according to our needs and interests. This authoring tool allows professors to create and/or link audiovisual learning pills, add textual explanations and design simple quizzes. Besides, professors can also tag, share and modify micro-content. Even though we have already mentioned complex standards for the characterisation of educational elements (such as LOM), we have decided to simplify this description as much as possible. To do this, we use sets of terms (bags of words) for more efficient management solutions based on simple natural language processing techniques.

The authoring tool was developed using the Django Web framework, a high-level Python Web framework based on an MVT (Model, View, Template) architecture pattern that encourages quick development and clean, pragmatic design. Additionally, we also have used the classic technologies for Web development (HTML5, JavaScript and CSS). Finally, we have decided to store micro-content in JSON format in a NoSQL database, such as MongoDB \cite{stevic2015enhancing}, to provide flexibility and the possibility to add different micro-learning templates in the near future. This approach allows different entries that may have different structures but still shape micro-content. This database (also shown as an SOA service in Fig. \ref{fig:cloud-platform})) can be queried from an HTTP API to retrieve the desired JSON of the specified micro-content.

\subsection{LTI: a solution to integrate micro-learning content in LMSs}
\label{sec:proposal-lti}

The full integration of micro-content in the LMS is key, because it guarantees that the LMS can exchange data with the micro-content. This enables the LMS to receive feedback (marks or academic results, for instance) that would enrich the student profile. We have checked different options for this to be possible. The first one would be to insert the HTML code directly; however, this is not easy for people with no technical profile and copy-pasting source code would block the Javascript scripts to connect to external services. The second one would be to upload the HTML file, which is easier than the previous solution, but it is not powerful enough to communicate with other systems. The third one would be to access the resource from a single URL, which is an end-user-friendly solution, but it does not support user identification or sending and receiving feedback from external platforms. Finally, we could create a SCORM package with the desired micro-learning content, but this requires another standard that increases the complexity of the system and the SCORM package does not allow interaction with third party services. We have found that none of these options is suitable for our purpose, since none provides both a complete interaction with the external micro-learning service and a user-friendly configuration.

After having considered the previous alternatives, we have decided to apply the {\em Learning Tools Interoperability} (LTI) standard of the IMS Global Learning Consortium \cite{LTIGuide}. This option allows the exchange of information between the micro-learning content and the LMS, which is essential to enrich the student profile with the results of the quizzes, for example. 

LTI technology has been widely used in many projects by several organizations and research groups in order to improve the learning experience and reuse the available resources. It is actually the leading standard to develop interoperable web tools in the e-learning environment. This feature is explained extensively in \cite{then2018competence} for the Moodle case, where both students and instructors can benefit from external resources. Several examples of LTI-compliant web tools developed for research and production purposes are shown in an old but widely referenced paper, \cite{severance2010ims}. Moreover, \cite{jurado2016ims} introduces a new software architecture for programming courses embedded in Moodle using LTI. On the other hand, some approaches have been done to enrich the learning environment in MOOCs using the LTI communication protocol. \cite{aleven2017integrating} shows an example where two external frameworks are attached to the edX platform to create an intelligent tutoring experience embedded in the courses \cite{aleven2018towards}. This option is clearly more flexible for extending the functionalities in the edX platform than using the native XBlock \cite{bakharia2017perspectivesx}. Furthermore, this technology has been applied as a free standard alternative to incorporate gamification activities in any LMS \cite{tuparov2018approaches} and exploiting the student data for learning analytics purposes \cite{perez2018standard}. Likewise, this solution has also been used to integrate sophisticated video games in an LMS and share user data with the other parties \cite{then2017introducing}. This way, the information can be processed and analyzed by the service provider \cite{then2018interfaces}.

LTI specification establishes a custom HTTP dialogue between the end-user, on behalf of their LMS or Tool Consumer (TC), and the Tool Provider (TP), who offers the web resources. Therefore, the TC acts as an intermediary, which locates and gets the connection parameters to obtain the LTI resources. Since the access to the TP must be protected from undesired sources, we can use the home institution of the student -which is known- to authenticate the user with the OAuth 1.0 protocol in the launching phase; this authentication will provide a session token to get access to the LTI tool. The micro-learning content can be integrated into the Moodle platform with the External Tool activity; this way, the professor only needs to configure a few parameters to use LTI. Besides, this solution could also be used in another LMSs compliant with the LTI architecture, like Blackboard, Canvas, Desire2Learn, Sakai, ANGEL, Jenzabar, PowerSchool (Haiku), Schoology or Etudes.

These flows are depicted in Fig. \ref{fig:cloud-platform}.
\begin{itemize}
\item On the one hand, the yellow and green flows show the process of updating micro-content in the LMS. First, the Administrator gives the Professor authorization to include micro-content in the LMS. The yellow flows shows (1) how the Administrator of the Cloud platform registers the LMS to which the Professor’s course belongs (ID of the LMS and shared secret) to authenticate the LTI protocol. Then, the green flow shows how the Professor (1) requires and (2) obtains the credentials (3) to create a new course activity in the LMS following the LTI specification from the micro-content URL.
\item On the other hand, the blue flow shows the process of a Student consuming a micro-content. The Student can (6) access the new micro-content through an LTI request (1) which is (2) provided by the LMS and is (3) sent to the LTI Tool Provider; the TP loads the micro-content from the HTTP API of the Micro-content Database and it is displayed to the Student in the web browser. 
\end{itemize}

The latest published version of LTI is the v1.3, but it is only available for some IMS members, so we have decided to develop the v1.1.1 instead of the v1.2 and v2.0 \cite{LTIOverview}, because it is public, stable and it would be easy to migrate to the latest one in the future \cite{LTIRoadmap}. The IMS also provides a detailed guide \cite{LTIRecipe} with all the necessary steps to implement a TP and several programming libraries in PHP or Java for easier development of LTI tools. Nevertheless, we have decided to develop our micro-content repository TP in Django/Python because it is easy to develop with the MVC framework.

The main reason for using LTI architecture was the possibility of exchanging information between the external content (micro-learning content in our case) and the LMS. More specifically, we are interested in collecting the results of the quiz embedded in the micro-learning content, as well as the interaction data (visualization of the visual content or interest on the micro-content, for instance). To do so, we have applied another IMS specification: the Learning Information Service (LIS) \cite{LISSpec}, specifically designed to send information from the TP, or a third party tool, to the TC. The LIS standard is a set of functionalities to exchange information -about users, courses, etc.- from the LMS with an external web service. More specifically, it is composed of up to six services, each one with its own Information Model and WSDL to correctly define the operations that can be performed. In the case of LTI, it includes some basic tasks from the Outcomes Management Service (OMS) related to the grading system of the LTI resource \cite{LTIGuide}. This way, we can access the grading system in LMS through XML messages, with the identifier of the micro-learning content (sourcedId), from the OMS server to the TC, which have been signed using OAuth1 body signing.

In Fig. \ref{fig:cloud-platform} the blue flow also shows the interaction with the LIS grading: the micro-content can have an embedded piece of code that (7)communicates to the LIS Grading service once the Student has completed the activity; it will then (8) upgrade the Student’s mark in the LMS.

\subsection{LTI and LIS: Implementation details}
\label{sec:proposal-lti-details}

\begin{figure}[htbp]
  \centering
  \includegraphics[width=0.7\textwidth]{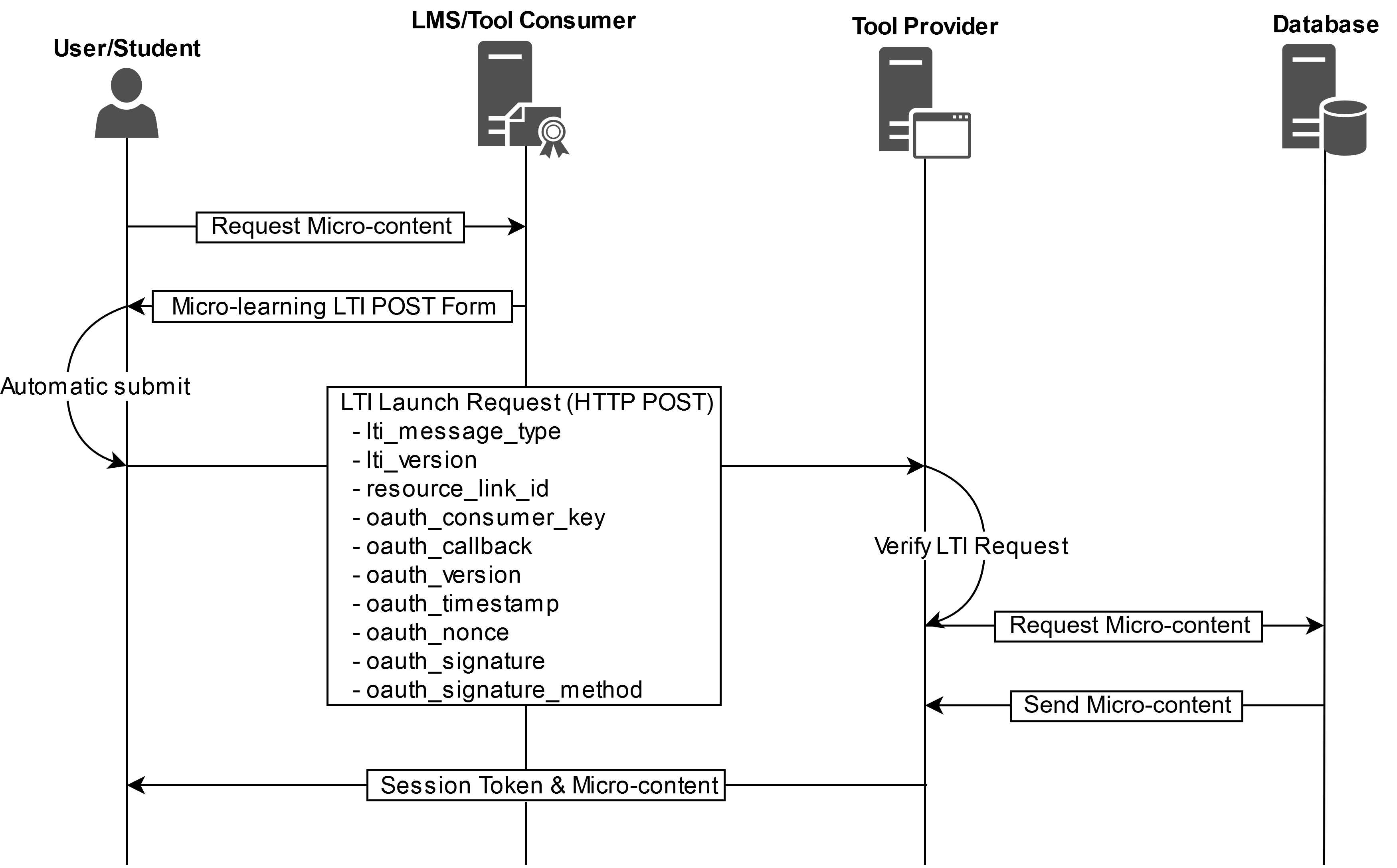}
  \caption{{\em Launching Phase} for accessing micro-learning content through LTI}
  \label{fig:LTI-launching}
\end{figure}

In this section, we describe the decisions taken to integrate the micro-learning content into Moodle through the LTI solution. The first stage is the {\em Launching Phase} shown in Fig. \ref{fig:LTI-launching}, which follows a series of steps. Initially, the student requests access to the micro-content and receives a POST form from the TC with all the required parameters (the OAuth and LTI fields) to access the desired LTI resource. This POST form is then automatically submitted using Javascript to send the launching HTTP message to the TP. This launch request HTTP POST contains the information shown in Fig. \ref{fig:LTI-launching}, which must be checked in order to validate or not the access. In brief, there are two checkpoints. On the one hand, the OAuth parameters are used to verify the signature ({\tt oauth\_signature}) according to the LMS identifier ({\tt oauth\_consumer\_key}) and password ({\tt oauth\_consumer\_key}) and other required parameters. On the other hand, some LTI parameters are also required, such as the message type ({\tt lti\_message\_type}), the version ({\tt lti\_version}), and the identifier of the resource within the LMS ({\tt resource\_link\_id}). We could also provide optional data to improve the experience in the TP, like the end-user identifier within the LMS ({\tt user\_id}) or the resource title ({\tt resource\_link\_title}). In fact, this second check-point that focuses on the LTI parameters is detailed in the pseudocode of the Listing below: 

\begin{lstlisting}[frame=single, language=Python, basicstyle=\tiny, caption={Pseudo-code of the LTI verification}, captionpos=b]
function verify_LTI(request, interval):
  ok = True
  ok = request.method == 'POST' && ok
  ok = request.lti_message_type == 'basic-lti-launch-request' && ok
  ok = request.lti_version == 'LTI-1p0' && ok
  ok = request.resource_link_id != '' && ok
  ok = is_registered(request.oauth_consumer_key) && ok
  ok = request.oauth_callback == 'about:blank'
  ok = request.oauth_version == '1.0' && ok
  ok = within_interval(request.oauth_timestamp, interval) && ok
  ok = is_unique(request.oauth_nonce, request.oauth_consumer_key) && ok
  secret = get_secret(request.oauth_consumer_key)
  signature = sign(request, request.oauth_consumer_key, secret,
  request.oauth_timestamp, request.oauth_nonce, request.oauth_signature_method)
  ok = request.oauth_signature == signature && ok
  if ok:
   	return (session_cookie, micro-content)
  else:
    return ERROR
%    \label{list:launching}
\end{lstlisting}

If the verification finds no problems, the end-user receives the desired resource and a session cookie that allows them to navigate over the TP. If instead the access is denied, that means the right TC credentials were not previously registered in the TP.

Once the {\em Launching phase} has finished, the TP works directly as an ordinary web service that requires a valid session cookie to access the micro-contents. Thus, this micro-content is displayed using HTML and Javascript technologies. The student can then visualize the text and video in the micro-content, answer the quiz and receive some feedback to their answers. In addition, the student only needs to be registered in his LMS and the LTI structure is transparent for him, that is, the micro-content resource seems to be embedded in the LMS as the authentication process is done by the LMS.

\begin{figure}[htbp]
  \centering
  \includegraphics[width=0.7\textwidth]{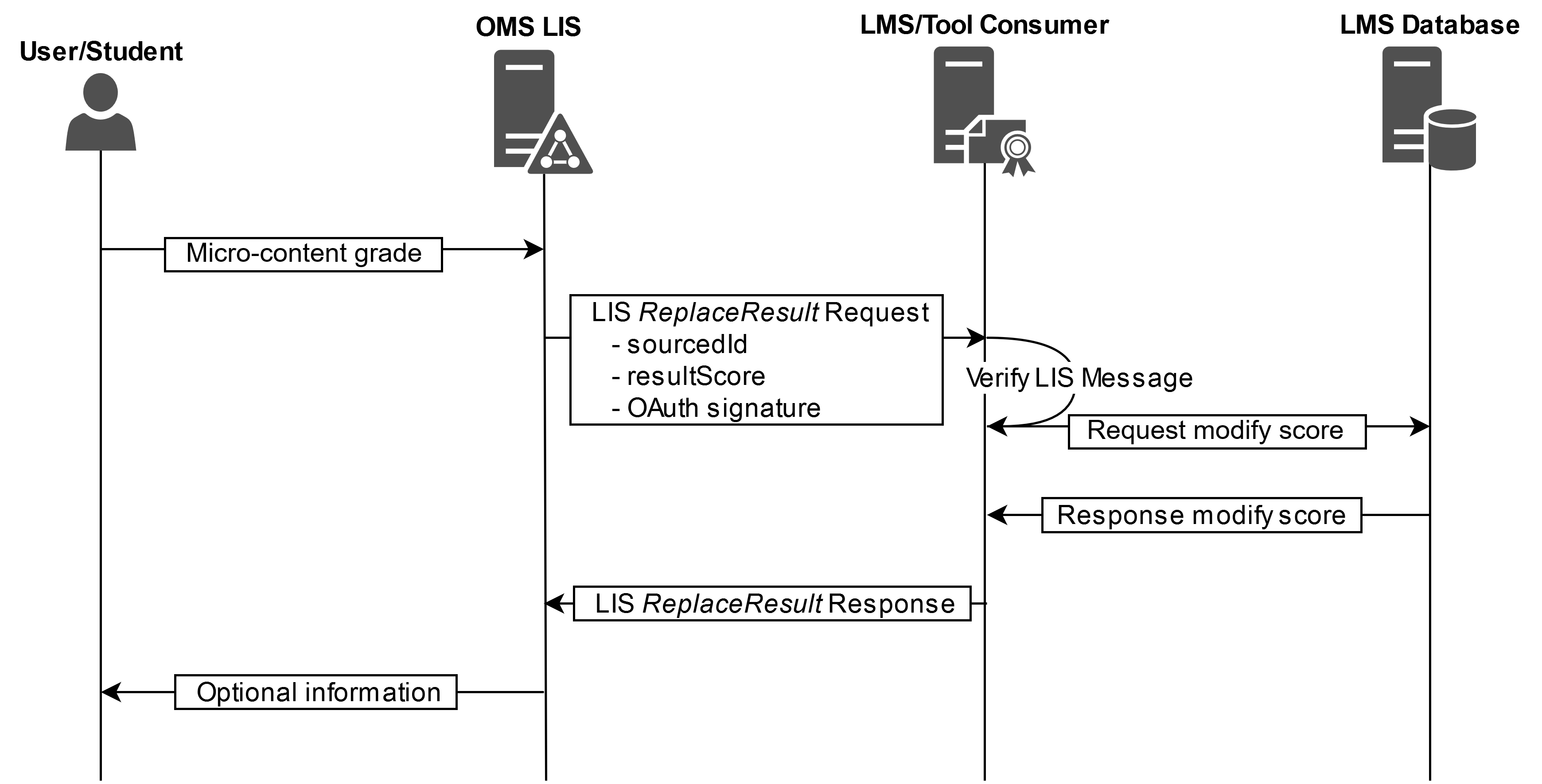}
  \caption{Diagram of the ReplaceResult mechanism for a micro-learning piece of content. The database symbol represents the information about the students that is already integrated in the LMS.}
  \label{fig:lis-replace-result}
\end{figure}

For the time being, the only implemented feature is the {\tt replaceResult} operation, which allows the modification of marks ({\tt resultScore}) within an interval of 0 to 1. This way, the student can complete the micro-learning content and answer the corresponding quiz from their usual LMS, and they can receive a result report while all the authentication and grading process is done by the LTI platform. The steps needed for this replacement process are shown in Fig. \ref{fig:lis-replace-result}. The student sends an HTTP request to the OMS LIS Grading service with the new grade (resultScore), as well as the session credentials (sourcedId and OAuth signature), associated to the correct LMS endpoint using the correct message format of the LIS specification. Then, the LMS verifies whether the LIS message is correct, if so, the response may contain some extra information that the student can see. Finally, this solution also supports the addition of more LIS services, allowing feedback from the micro-contents in the LMS and data analysis for the professor to check the effectiveness of their resources.

\begin{figure}[htbp]
  \centering
  \includegraphics[width=0.6\textwidth]{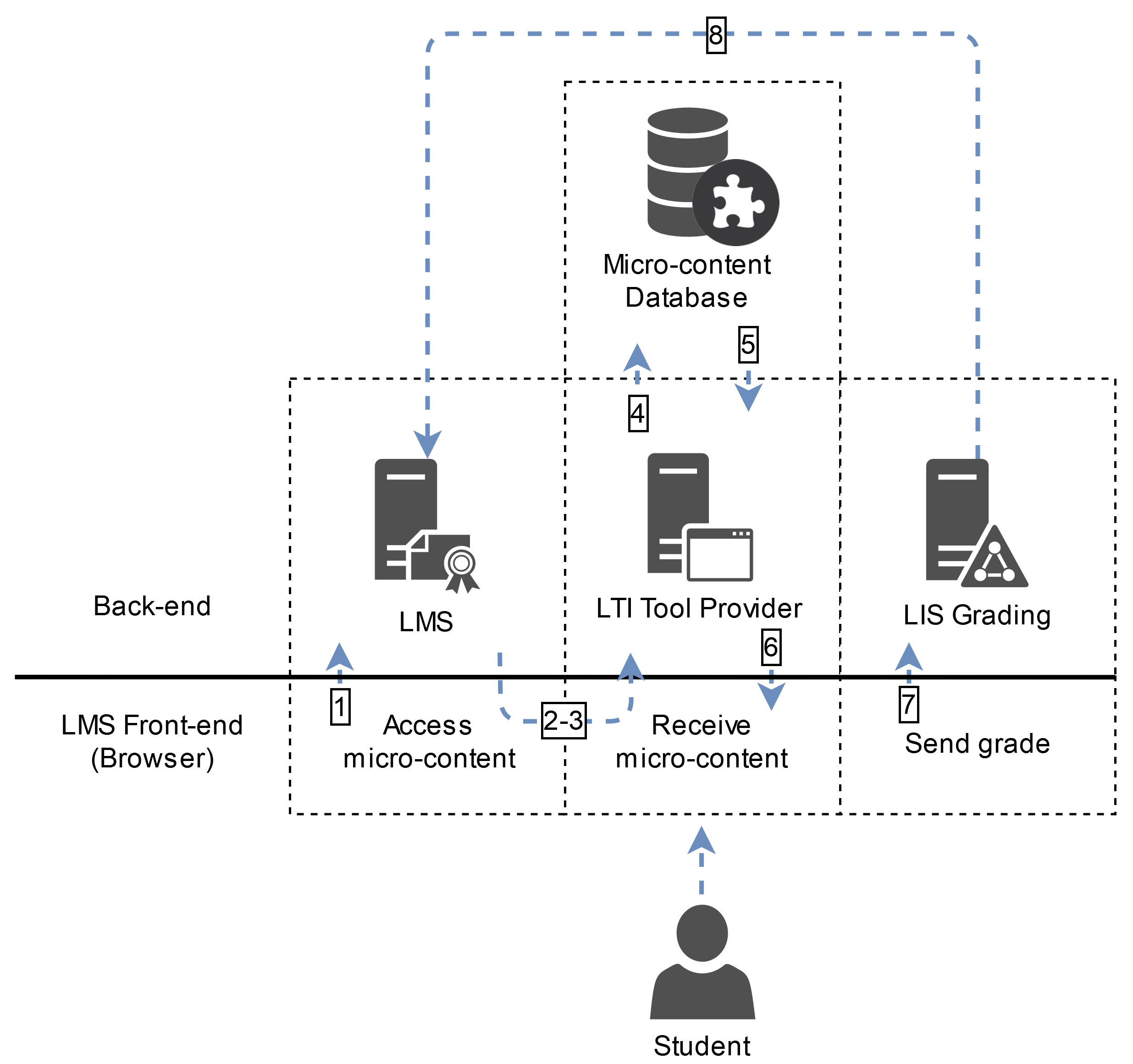}
  \caption{Flow to interact with a micro-content within the LMS}
  \label{fig:lms-micro-content-flow}
\end{figure}

To conclude, the interaction between the student and the LMS and the micro-content included is summarized in Fig. \ref{fig:lms-micro-content-flow}. In particular, the student only has access to the LMS Front-end (web browser) to perform the whole path. This means that, from the beginning of the student's access to the micro-content activity, they receive (1) all the necessary information from the LMS in order to communicate with the pieces of our Cloud platform in a transparent way: they receive the LTI POST message from the LMS, which has to be sent automatically to the Tool Provider (2,3) to receive the loaded micro-content (6); then, the Student does the micro-content quiz and sends the achieved mark to the LIS service (7) according to the received endpoint in the previous step.

\section{Evaluation}
\label{sec:evaluation}

With the aim of assessing the opinion about micro-learning and our approach, we have prepared a survey form for experts in distance learning and professors at Higher education institutions. We have received feedback from 58 participants from two different networks. On the one hand, the Spanish Network of Learning Analytics (SNOLA\footnote{https://snola.es/}). This network brings together researchers and professors from different universities in Spain. On the other hand, we also sent the questionnaire to professors and researchers with in the ELEMEND CBHE project\footnote{http://elemend.ba/}. This is a international project funded by the EU that pursues updating the electrical engineering curricula (BSc and MSc levels) in Western Balkan countries. Within this project, new distance learning paradigms, like micro-learning, have been taking into account to develop new content for students and workers in electrical energy markets. 

\begin{figure}[htbp]
  \centering
  \includegraphics[width=1 \textwidth]{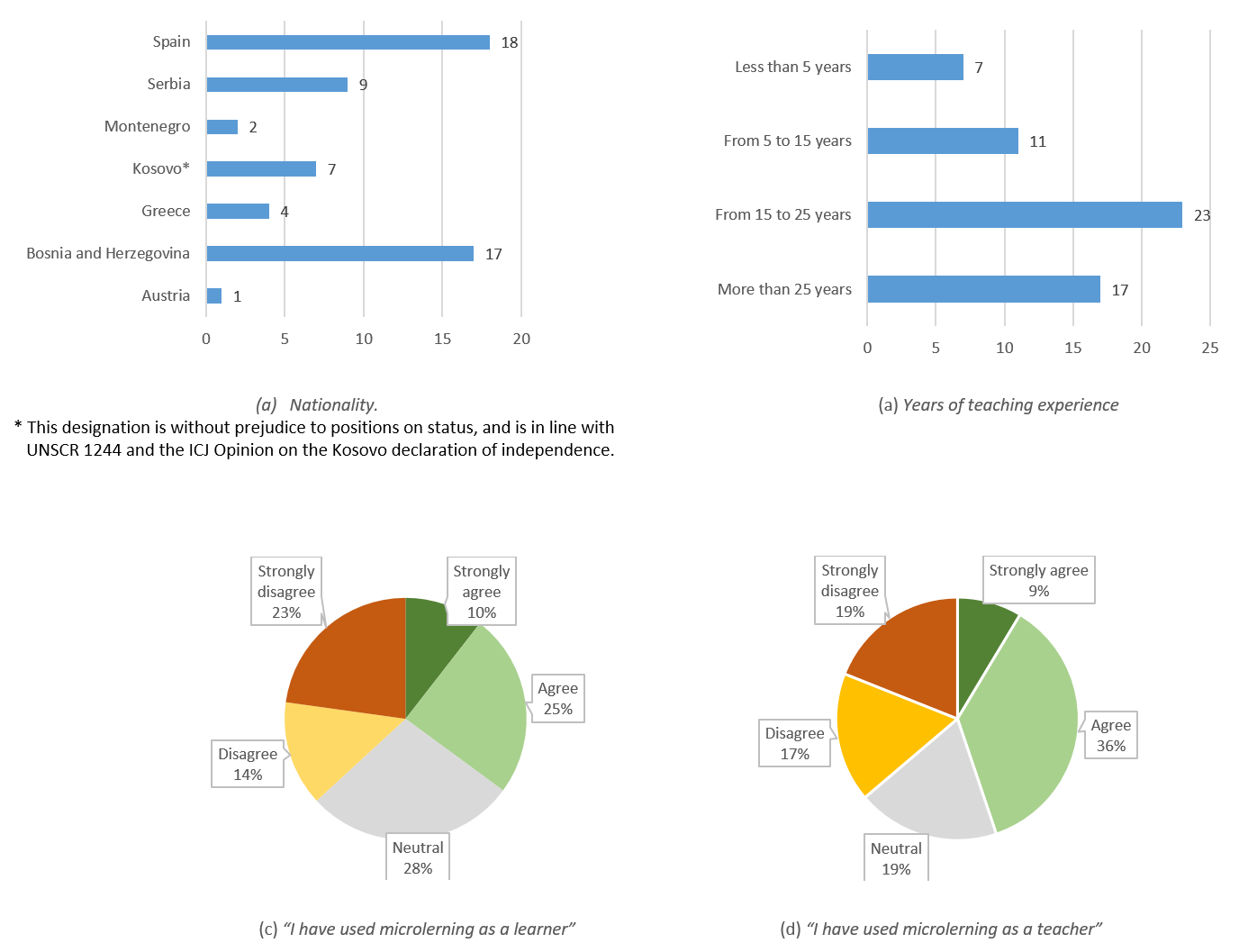}
  \caption{Information about the participants in the survey}
  \label{fig:surveydemo}
\end{figure}

The survey was divided into three parts. The first one to obtain information about the expert (age range, nationality and teaching experience). According to the results (Fig. \ref{fig:surveydemo}), the majority of the participants have more than 15 years of teaching experience (69\%) and from two Europan regions: Spain (31\%) and Western Balkans (29\%). Only 35\% of them have experienced micro-learning as learners and a slightly higher percentage (45\%) has experience micro-learning as teacher.

\begin{figure}[htbp]
  \centering
  \includegraphics[width=1 \textwidth]{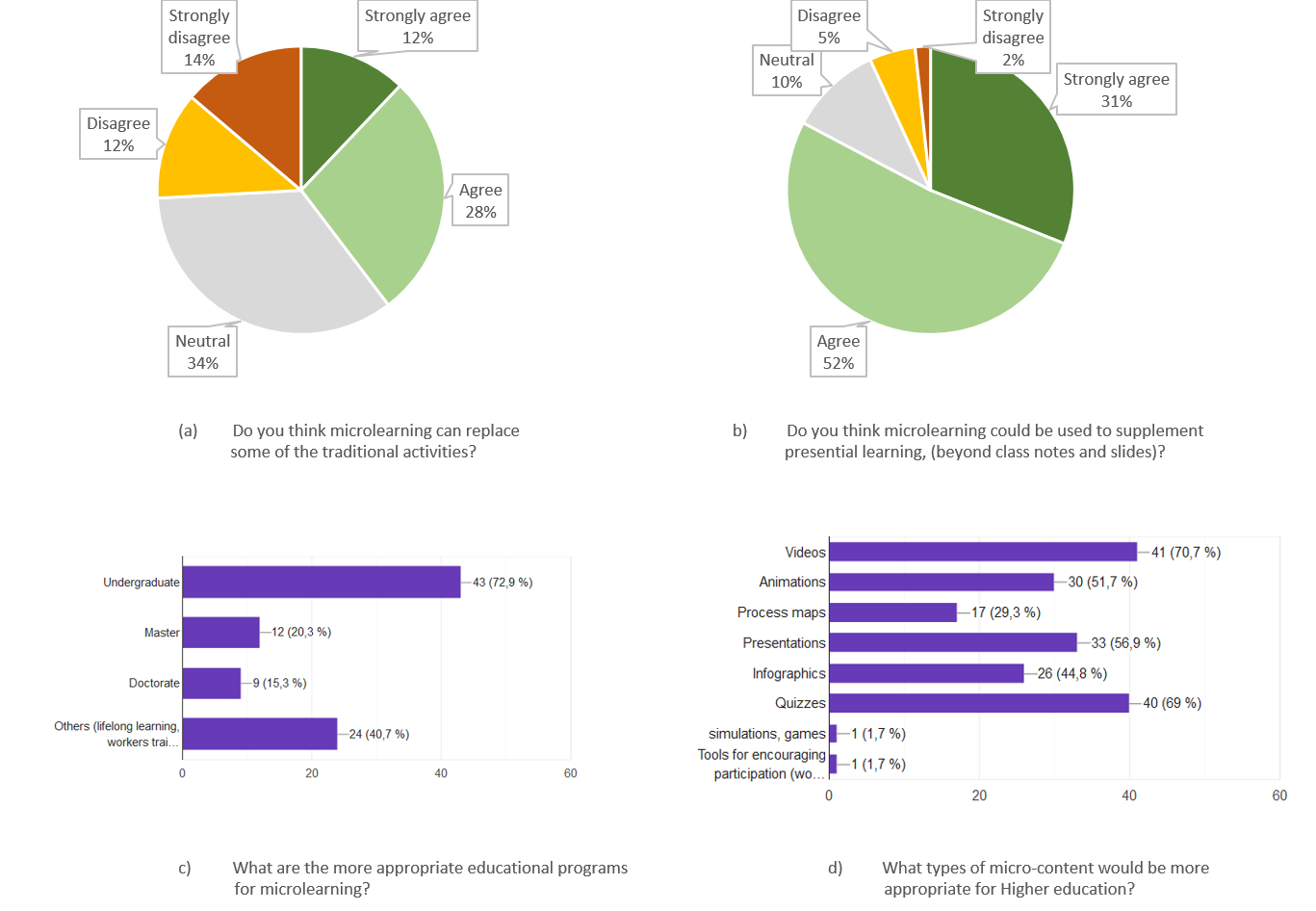}
  \caption{General opinion about micro-learning}
  \label{fig:microlearningsurvey}
\end{figure}

In the second part, we asked about the advantages and/or disadvantages of the micro-learning paradigm (Fig \ref{fig:microlearningsurvey}). The opinion is balance about the possibility of replacing some traditional teaching activities by micro-learning ones: 40\% agree, but 28\% disagree. However, a great majority thinks that micro-learning is a good element to supplement face-to-face learning: 83\% agrees, whereas only a 7\% do not share this opinion. The results show that the majority of the participant consider that micro-learning may be useful for undergraduate and lifelong learning and corporate training, but its suitability for master and doctorate is not supported (less than a one fifth consider that a good approach).

\begin{figure}[htbp]
  \centering
  \includegraphics[width=1 \textwidth]{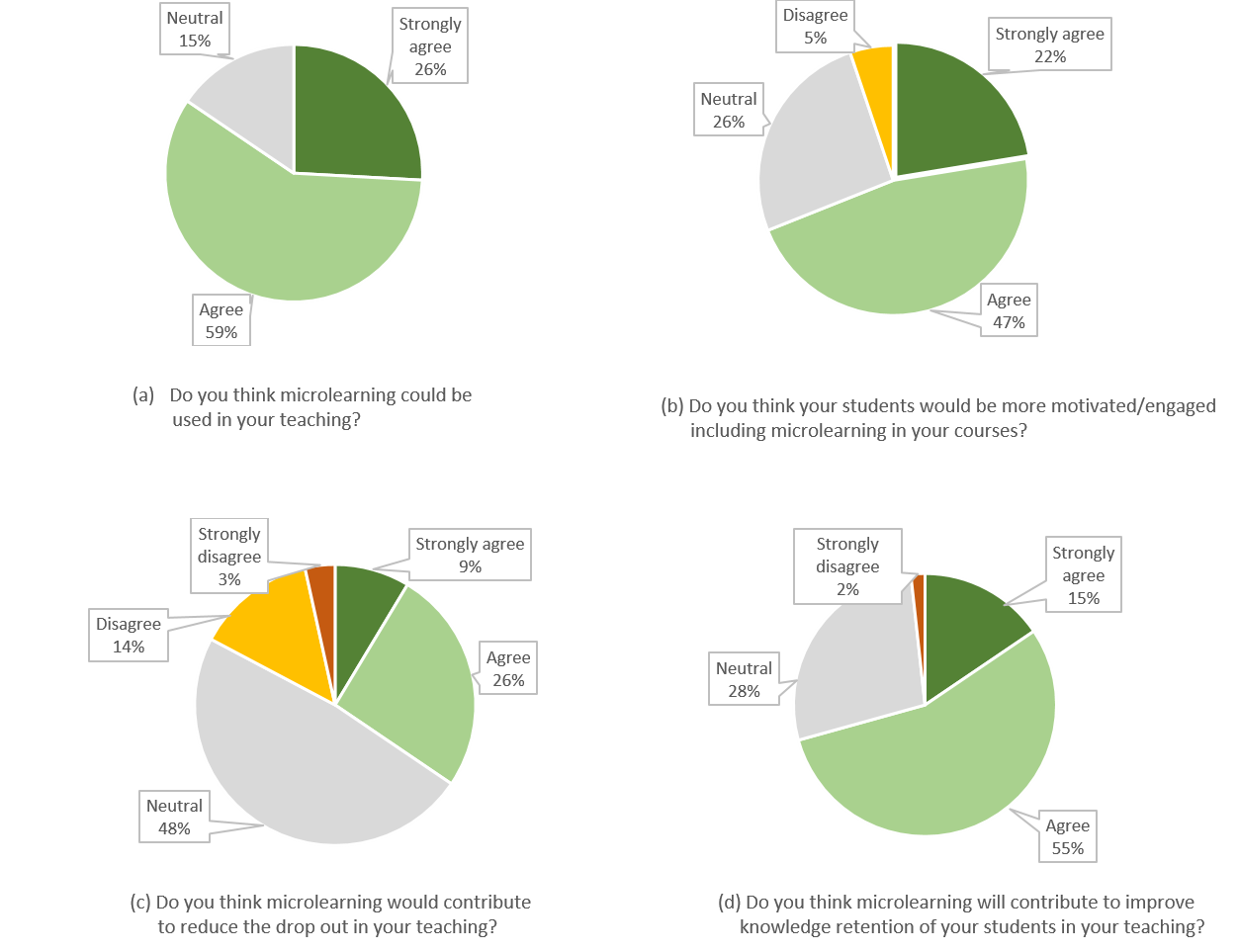}
  \caption{Micro-learning in your courses}
  \label{fig:Survey3}
\end{figure}

Finally, we have explicitly asked about the application of micro-learning in their courses (Fig. \ref{fig:Survey3}). A large majority (84.7\%)  consider that the micro-learning paradigm could be used for their teaching activity and 72.8\% consider micro-learning is even appropriate for their students. The majority of the participants consider micro-learning would contribute to improve the engagement (67.7\%) and knowledge retention (71.2\%). The survey included other questions about the main goals and barriers of introducing micro-learning into their teaching. Regarding this, there is not a wide agreement about the micro-learning contribution to reduce the drop-out (34\% in favour and 50\% neutral). A good percentage of participants (60.3 \%) would use micro-learning at the end of the lessons to obtain feedback about the new content and to reinforce the learning process. The main motivations to use the new paradigm would be: (i) to improve students motivation (50.9\%); (ii) to improve the students engagement (64.9\%); and (iii) to facilitate students self-assessment (49.1\%). However, the participants consider that the main barrier to adopt this new paradigm is the preparation of new content, images, videos, interactive content, etc., (65.5\%). However, the great majority considers there are not relevant barriers to deliver micro-content to students, only 8.8\% disagrees.

\section{Discussion}
\label{sec:discussion}

The micro-learning philosophy of breaking down new concepts into small pills of content to be consumed in short periods of time has emerged as a positive approach for workers. They are usually overwhelmed by their jobs and the need for being constantly up-to-date and this paradigm has clear advantages for the trainees. First, it improves the level of engagement with the educational experience, since micro-learning exclusively focuses on the transmission of relevant information, with the aid of short visual and interactive elements. As a consequence, the possibility of falling behind decreases, and the student is less likely to drop out or tune out during the training. Second, it improves retention, since micro-learning activities are short, concise and coherent with short-term memory, which is estimated to be capable of managing no more than 4 elements simultaneously. In fact, the micro-learning paradigm fits in really well with the learning techniques in accordance with the Forgetting curve, based on three main pillars: (i) spaced repetition, with an adequate frequency to facilitate the cognitive process; (ii) variety of formats and contexts for the same concept or element and (iii) the interleaving of different concepts or elements to be learnt so as to facilitate the transfer and the acquisition of new knowledge. Finally, it improves the environment of the process, since the content is adapted to the learners’ schedule and their levels of education and prior knowledge. The learning environment is more flexible and effective. Furthermore, practice distributed over time is known to increase the level of efficiency in the learning process (according to some studies, by 17\%).

Additionally, it is also a paradigm that benefits trainers in at least two ways. First, it allows for a quicker, cheaper and simpler learning process. Learning costs are estimated to drop by 50\%, tripling the speed at which new content is assimilated. Since micro-content is sequenced in a modular manner, it is easier to manage with the aid of a Content Management System (CMS). Besides, it is also easier to adapt short amounts of content to a range of contexts, such as translations into different languages. Second, it allows access to training as needed (just-in-time): training is offered and consumed when demanded. Since content is clear and concise, it can be created at a lower cost, which allows for greater agility when generating new content tailored to the rapidly changing needs of corporate environments.

However, this training paradigm should not be considered the solution to all teaching problems, including those related to distance learning. Despite its clear advantages, this paradigm also presents an important shortcoming. Micro-learning does not perform so well when teaching complex, abstract concepts that require the combination of multidisciplinary knowledge. In these cases, trainees need to spend a significant amount of time to guarantee the comprehension of the new concepts, to ensure that these are appropriately linked to those previously acquired and to put them into practice to reinforce the cognitive processes and guarantee their permanence. In other words, micro-learning is a very adequate option for simple, rote learning, which can be either combined with more complex training or aimed at the acquisition of basic skills.  

In our analysis, we detected that university lectures have a favourable position for the adoption of this new approach because of the following reasons: increase of motivation and engagement of student as well as improvement in the retention levels. This is why the majority considers it appropriate to include micro learning content at the end of the lessons to reinforce the learning process and to receive feedback. Besides, the integration in LMSs is not considered a problem, on the contrary, LMSs make it easy to deliver this new content to students.

\section{Conclusions and further work}
\label{sec:conclusion}

This paper summarizes the main approaches in this field in recent years and introduces a new solution to integrate micro-learning into traditional distance learning platforms. This hybrid approach combines the advantages of the micro-learning paradigm and the benefits of the traditional solutions, enabling the introduction of micro-content, not only for simple knowledge, but also to reinforce the acquisition of more complex skills. Students enjoy the attractive and easy-to-consume micro-content, while trainers can combine both approaches in the same platform. Our solution is based on a Service-Oriented Architecture of different services that are deployed in the cloud. In order to guarantee the full integration of the micro-content in a traditional LMS, we propose using LTI and LIS as the two standards that allow the micro-content to exchange data with the LMS and register the student's feedback on their profile.

In order to assess our proposal, we have prepared a survey for lectures from different countries with experience in distance learning. The participants belong to three networks that bring together specialists in learning analytics and distance learning, as well as lectures from Western Balkan countries that participate in the ELEMEND CBHE project. In this project, micro-learning is being used according to our proposal: including micro-learning content within traditional LMS (Moodle). Most of the participants consider this approach is convenient as a supplement tool in higher education and that it is useful for their courses. Although there is no agreement in its potential for reducing the drop-out, a large majority consider it is a good proposal to reinforce the teaching process as well as to improve the engagement of students. The main drawback is the effort needed to create audiovisual and interactive content for this new paradigm.

Since one of the relevant aspects of micro-learning is the active participation of learners in the process of co-creation and distribution of micro-content \cite{Linder2006}, we are currently working on providing the mechanisms to allow both students and professors to have a more collaborative role in the creation and learning processes. Finally, we are also analyzing the possibilities of combining bots and an existent talk-application to distribute and execute the micro-content, with the aim to reach students in their free time.

\section*{Acknowledgements}
\noindent This work is funded by: the European Regional Development Fund (ERDF) and the Galician Regional Government under the agreement for funding the Atlantic Research Center for Information and Communication Technologies (AtlantTIC), the Spanish Ministry of Economy and Competitiveness under the National Science Program (TEC2017-84197-C4-2-R) and the Education, Audiovisual and Culture Executive Agency under the Capacity Building Programme with the project ELEMEND (585681-EPP-1-2017-1-EL-EPPKA2-CBHE-JP).



\end{document}